\documentclass[11pt,draftcls,onecolumn,journal]{IEEEtran}
\usepackage{amsmath}
\usepackage{amssymb}
\usepackage{enumerate}
\usepackage{theorem}
\usepackage{graphicx,color}
\usepackage{pifont}
\usepackage[noadjust]{cite}
\usepackage{citesort}
\usepackage[english]{algorithme}

\def\trans{\mbox{\tiny $\top$}}
\newcommand{\vect}[1]{{\boldsymbol #1}}    
\newcommand{\RR}{{\ensuremath{\mathbb R}}}

\newcommand{\NN}{{\ensuremath{\mathbb N}}}

\title{A Hierarchical Bayesian Model for Frame Representation}
\graphicspath{{./Figures/}}
\author {L. Cha\^ari, \textit{Student Member, IEEE}, J.-C. Pesquet, \textit{Senior Member, IEEE}, J.-Y. Tourneret,\\ 
\textit{Senior Member, IEEE}, Ph. Ciuciu, \textit{Member, IEEE} and A. Benazza-Benyahia, \textit{Member, IEEE}
\thanks{L. Cha\^ari and J-C Pesquet are with LIGM and UMR-CNRS 8049, Universit{{\'e}} Paris-Est, Champs-sur-Marne, 77454 Marne-la-Vall{\'e}e, France.
E-mail: $\{$lotfi.chaari,jean-christophe.pesquet$\}$@univ-paris-est.fr, J.-Y. Tourneret is with
the University of Toulouse, IRIT/ENSEEIHT/TSA,
31071 Toulouse, France. E-mail: jean-yves.tourneret@enseeiht.fr.
Ph. Ciuciu is with CEA/DSV/$\mathrm{I}^2$BM/Neurospin, CEA Saclay,
Bat. 145, Point Courrier 156, 91191 Gif-sur-Yvette cedex, France.
E-mail: philippe.ciuciu@cea.fr, A. Benazza-Benyahia is with the
Ecole Sup\'erieure des Communications de Tunis (SUP'COM-Tunis), Unit\'e de Recherche en Imagerie Satellitaire et ses Applications (URISA) 
, Cit\'e Technologique des Communications, 2083, Tunisia.
E-mail: benazza.amel@supcom.rnu.tn }
\thanks{This work was supported by grants from R\'egion Ile de France and the Agence Nationale de la Recherche under grant
ANR-05-MMSA-0014-01.}}

\newcommand{\scal}[2]{\left\langle{{#1}|{#2}}\right\rangle}

\newtheorem{theorem}{Theorem}[section]

\theoremstyle{plain}{\theorembodyfont{\rmfamily}%
}

\renewcommand\theenumi{(\roman{enumi})}
\renewcommand\theenumii{(\alph{enumii})}

\begin{document}
\maketitle
\begin{abstract}
In many signal processing problems, it may be fruitful to represent
the signal under study in a frame. If a probabilistic approach is
adopted, it becomes then necessary to estimate the hyper-parameters
characterizing the probability distribution of the frame
coefficients. This problem is difficult since in general the frame
synthesis operator is not bijective. Consequently, the frame
coefficients are not directly observable. This paper introduces a
hierarchical Bayesian model for frame representation. The posterior
distribution of the frame coefficients and model hyper-parameters is
derived. Hybrid Markov Chain Monte Carlo algorithms are subsequently
proposed to sample from this posterior distribution. The generated
samples are then exploited to estimate the hyper-parameters and the
frame coefficients of the target signal. Validation experiments show
that the proposed algorithms provide an accurate estimation of the frame coefficients and hyper-parameters.
Application to practical problems of image denoising show the
impact of the resulting Bayesian estimation on the
recovered signal quality.

\end{abstract}

\begin{keywords}
Frame representations, Bayesian estimation, MCMC, Gibbs sampler, Metropolis Hastings, hyper-parameter estimation, Generalized Gaussian, sparsity, compressed sensing, wavelets.
\end{keywords}

\newpage
\section{Introduction}\label{sec:intro}
Data representation is a crucial operation in many signal and image
processing applications. These applications include signal and image
reconstruction \cite{kawahara_93,chaari_08} , restoration
\cite{miller_99,chaux_05} and compression
\cite{martin_01,meuleneire_08}. In this respect, many linear
transforms have been proposed in order to obtain suitable signal
representations in other domains than the original spatial or
temporal ones. The traditional Fourier and discrete cosine
transforms provide a good frequency localization, but at the expense
of a poor spatial or temporal localization. To improve localization
both in the spatial/temporal and frequency domains, the wavelet
transform (WT) was introduced as a powerful tool in the $1980$'s
\cite{Mallat_S_98}. Many wavelet-like basis decompositions have been
subsequently proposed offering different features. For instance, we
can mention the wavelet packets \cite{Coi_92} or the grouplet bases
\cite{Mallat_08}. To further improve signal representations,
redundant linear decomposition families called \emph{frames} have
become the focus of many works during the last decade. For the sake of clarity, it must be pointed out that the term frame \cite{Candes_02} is understood in the sense of Hilbert space theory and not in the sense of some recent works like \cite{Destrempes_06}.\\
The main advantage of frames lies in their flexibility to capture
local features of the signal. Hence, they may result in sparser
representations as shown in the literature on curvelets
\cite{Candes_02}, bandelets \cite{Pennec_05} or dual-trees
\cite{Chaux_06} in image processing. However, a major difficulty
when using frame representations in a statistical framework is to
estimate the parameters of the frame coefficient probability
distribution. Actually, since frame synthesis operators are
generally not injective, even if the signal is perfectly known, the
determination of its frame coefficients is an underdetermined
problem.

This paper studies a hierarchical Bayesian approach to estimate the
frame coefficients and their hyper-parameters. Although this approach is
conceptually able to deal with any desirable distribution for the frame
coefficients, we focus in this paper on generalized Gaussian (GG) priors.
Note however that we do not restrict our attention to log-concave GG prior probability density functions (pdf),
which may be limited for providing accurate models of sparse signals \cite{seeger_07}.
In addition, the proposed
method can be applied to noisy data when imprecise measurements of
the signal are only available. Our work takes advantage of the
current developments in Markov Chain Monte Carlo (MCMC) algorithms
\cite{Robert_04,Cappe_02,Andrieu_01} that have already been investigated for instance in image separation \cite{Ichir_03}, image restoration \cite{Jalobeanu_02} and brain activity detection in functional MRI \cite{Makni_05,Makni_08}. These algorithms have also been investigated for signal/image processing problems with sparsity constraints. These constraints may be imposed in the original space like in \cite{dobigeon_09}, where a sparse image reconstruction problem is assessed in the image domain. They may also be imposed on some redundant representation of the signal like in \cite{blumensath_07}, where a time-series sparse coding problem is addressed.\\

Hybrid MCMC algorithms
\cite{Zeger_91,Tierney_94} are designed combining Metropolis-Hastings
(MH) \cite{Hastings_70} and Gibbs \cite{Geman_84} moves to sample
according to the posterior distribution of interest. MCMC algorithms
and WT have been jointly investigated in some works dealing with
signal denoising in a Bayesian framework
\cite{Leporini_01,Muller_99,Heurta_05,Ichir_03}.
However, in contrast with the present work where overcomplete frame representations are considered, these works are limited to wavelet
bases for which the hyper-parameter estimation problem is much easier to handle.\\
This paper is organized as follows. Section \ref{sec:background}
presents a brief overview on the concepts of frame and frame
representation. The hierarchical Bayesian model proposed for frame
representation is introduced in Section \ref{sec:bayes}. Two algorithms for
sampling the posterior distribution are proposed in Section \ref{sec:sampling}.
To illustrate the effectiveness of these algorithms, experiments on both
synthetic and real world data are presented in Section
\ref{sec:simuls}. In this section, applications to image recovery problems are also considered. Finally some conclusions are drawn in Section~\ref{sec:conclusion}.

\section{Problem Formulation}\label{sec:background}
\subsection{The frame concept} \label{sec:FC}
In the following, we will consider real-valued digital signals of
length $L$ as elements of the Euclidean space $\mathbb{R}^{L}$
endowed with the usual scalar product and norm denoted as
$\scal{.}{.}$ and $\Vert \cdot \Vert$, respectively. Let $K$ be an
integer greater than or equal to $L$. A family of vectors
$(\vect{e}_k)_{1\le k\le K}$ in the finite-dimensional space
$\mathbb{R}^L$ is a frame when there exists a constant $\mu$ in
$]0,+\infty[$ such that\footnote{The classical upper bound condition
is always satisfied in finite dimension. In this case, the frame
condition is also equivalent to saying that the frame operator has full rank $L$.}
\begin{equation} \label{eq:inequality}
\forall \vect{y}\in \mathbb{R}^L, \qquad \mu \|\vect{y}\|^2 \leq \sum_{k=1}^K |\scal{\vect{y}}{\vect{e}_k}|^2.\; 
\end{equation}
If the inequality \eqref{eq:inequality} becomes an equality,
$(\vect{e}_k)_{1\le k\le K}$ is called a \textit{tight} frame. The
bounded linear frame analysis operator $F$ and the adjoint synthesis
frame operator $F^* $ are defined as
\begin{align}
F\colon \mathbb{R}^L  &\to \RR^K \\ \nonumber \vect{y} &\mapsto
(\scal{\vect{y}}{\vect{e}_k})_{1\le k \le K}, \\
F^*  \colon \RR^K   &\to \mathbb{R}^L \\ \nonumber (\xi_k)_{1\le k
\le K} &\mapsto \sum_{k=1}^K \xi_k \vect{e}_k.
\end{align}
Note that $F$ is injective whereas $F^*$ is surjective. When
$F^{-1}=F^*$, $(\vect{e}_k)_{k\in \mathbb{K}}$ is an orthonormal
basis. A simple example of a redundant frame is the union of $M > 1$
orthonormal bases. In this case, the frame is tight with $\mu = M$
and thus, we have $F^*F=M \mathrm{I}$ where $\mathrm{I}$ is the
identity operator.

\subsection{Frame representation}\label{sec:FD}
An observed signal $\vect{y} \in \RR^L$ can be written according to
its frame representation (FR) involving coefficients $\vect{x} \in
\RR^K$ as follows
\begin{equation}\label{eq:delta}
\vect{y}= F^*\vect{x} + \vect{n}
\end{equation}
where $\vect{n}$ is the error between the observed signal $\vect{y}$ and its FR $F^*\vect{x}$. 
This error is modeled by imposing that $\vect{x}$ belongs to the
closed convex set
\begin{equation}\label{eq:Cdelta}
C_{\delta}=\{\vect{x}\in \mathbb{R}^K \mid N(\vect{y}-F^*\vect{x})
\leq \delta \}
\end{equation}
where $\delta\in [0,\infty[$ is some error bound and $N(.)$ can be any norm on $\RR^L$.\\
In signal/image recovery problems, $\vect{n}$ is nothing but an additive noise that corrupts the measured data. By adopting a probabilistic approach,
$\vect{y}$ and $\vect{x}$ are assumed to be realizations of random
vectors $\vect{Y}$ and $\vect{X}$. In this context, our goal is to
characterize the probability distribution of $\vect{X}|\vect{Y}$, by
considering some parametric probabilistic model and by estimating
the associated hyper-parameters.\\
A useful example where this characterization may be of great
interest is frame-based signal/image denoising in a Bayesian
framework. Actually, denoising in the wavelet domain using wavelet
frame decompositions has already been investigated since the seminal work \cite{Donoho_denoising_95} as this kind of representation
provides sparse description of regular signals. The related
hyper-parameters have then to be estimated.

When $F$ is bijective and $\delta = 0$, this estimation can be
performed by inverting the transform so as to deduce $\vect{x}$ from
$\vect{y}$ and by resorting to standard estimation techniques on
$\vect{x}$. However, as mentioned in Section \ref{sec:FC}, for
redundant frames, $F^*$ is not bijective, which makes the
hyper-parameter estimation problem more difficult since deducing $\vect{x}$ from
$\vect{y}$ is no longer unique. This paper
presents hierarchical Bayesian algorithms to address this issue.

\section{Hierarchical Bayesian Model}\label{sec:bayes} In a Bayesian
framework, we first need to define prior distributions for the frame
coefficients. For instance, this prior may be chosen so as to
promote the sparsity of the representation. In the following,
$f(\vect{x} |\vect{\theta})$ denotes the pdf of the frame coefficients that depends on an unknown
hyper-parameter vector $\vect{\theta}$ and $f(\vect{\theta})$ is the
a priori pdf of the hyper-parameter vector $\vect{\theta}$. In
compliance with the observation model~\eqref{eq:delta}, $f({\vect{y}
| \vect{x}})$ is a uniform distribution on the closed convex set
$D_{\delta}$ defined as
\begin{equation}
D_{\delta}=\{\vect{y}\in \mathbb{R}^L  \mid N(\vect{y}-F^*\vect{x})
\leq \delta \}
\end{equation}
where $\delta > 0$. Denoting by $\vect{\Theta}$ the random variable
associated with the hyper-parameter vector $\vect{\theta}$ and using
the hierarchical structure between $\vect{Y},\vect{X}$ and
$\vect{\Theta}$, the conditional distribution of
$(\vect{X},\vect{\Theta})$ given $\vect{Y}$ can be written as
\begin{equation} \label{eq:posterior}
f(\vect{x},\vect{\theta}|\vect{y}) \propto f(\vect{y}|\vect{x})
f(\vect{x}|\vect{\theta})f(\vect{\theta})
\end{equation}
where $\propto$ means \textit{proportional to}.

In this work, we assume that frame coefficients are a priori independent with
marginal GG distributions. This assumption
has been successfully used in many studies
\cite{Mallat_89,Antonini_92,Joshi_97,Simoncelli_96,Moulin_98,Do_02} and leads to
the following frame coefficient prior
\begin{equation}
f(x_k | \alpha_k, \beta_k)=\frac{\beta_k}{2 \alpha_k \Gamma(1/\beta_k)}
\exp \left( - \frac{|x_k|^{\beta_k}}{\alpha_k^{\beta_k}}
  \right)
\end{equation}
where $\alpha_k > 0 ,\beta_k > 0$ (with $k \in \{1,\ldots,K\}$) are
the scale and shape parameters associated with $x_k$, which is the
$k$th component of the frame coefficient vector $\vect{x}$ and
$\Gamma(.)$ is the Gamma function. Note that small values of the shape
parameters are appropriate for modelling sparse signals. When $\forall k \in \{1,\ldots,K\}$, $\beta_k =1$, a Laplace prior is
obtained, which was shown to play a central role in sparse signal
recovery \cite{Seeger_08} and compressed sensing \cite{Babacan_09}.

By introducing $\forall k \in \{1,\ldots,K\}$, $\gamma_k=\alpha_k^{\beta_k}$, the frame prior can be
rewritten as\footnote{The interest of this new parameterization will be
clarified in Section \ref{sec:sampling}.}
\begin{equation}
\label{eq:priorreparam}
f(x_k | \gamma_k, \beta_k)= \frac{\beta_k}{2 \gamma_k^{1/\beta_k}
\Gamma(1/\beta_k)} \exp \left( - \frac{|x_k|^{\beta_k}}{\gamma_k}
  \right).
\end{equation}
The distribution of a frame coefficient generally differs from one
coefficient to another. 
However, some frame coefficients can have very
similar distributions (that can be defined by the same
hyper-parameters $\beta_k$ and $\gamma_k$). As a consequence, we
propose to split the frame coefficients into $G$ different groups.
The $g$th group will be parameterized by a unique
hyper-parameter vector denoted as
$\vect{\theta_g}=(\beta_g,\gamma_g)$ (after the
reparameterization mentioned above). In this case, the frame prior
can be expressed as
\begin{equation}
f(\vect{x}|\vect{\theta}) = \prod_{g=1}^G \left[ \left(
\frac{\beta_g}{2 \gamma_g^{1/\beta_g} \Gamma(1/\beta_g)}
\right)^{n_g}\exp \left( - \frac{1}{\gamma_g} \sum_{k\in S_g} |x_k|^{\beta_g}
  \right) \right]
\end{equation}
where the summation covers the index set $S_g$ of the elements of
the $g$th group containing $n_g$ elements and
$\vect{\theta}=(\vect{\theta_1},\ldots,\vect{\theta_G})$. Note that in our simulations, each group $g$ will correspond to a given wavelet subband. A coarser classification may be made when using multiscale frame representations by considering that all the frame coefficients at a given resolution level belong to a same group.

The hierarchical Bayesian model for the frame decomposition is
completed by the following improper hyperprior
\begin{align}
f(\vect{\theta}) & = \prod_{g=1}^G f(\vect{\theta_g}) \nonumber\\
&= \prod_{g=1}^G \left[ f(\gamma_g) f(\beta_g) \right] \nonumber\\
&\propto \prod_{g=1}^G  \left[\frac{1}{\gamma_g}
\mathsf{1}_{\mathbb{R}^+}(\gamma_g)\mathsf{1}_{[0,3]}(\beta_g)
\right]
\end{align}
where for a set $A \subset \RR$,
\begin{align}
\mathsf{1}_A(\xi)= &
\begin{cases}
1 & \mbox{if $\xi \in A$}\\
0 & \mbox{otherwise.}
\end{cases}
\end{align}
The motivations for using this kind of prior are summarized
below:
\begin{itemize}
\item the interval $[0,3]$ covers all possible values of
$\beta_g$ encountered in practical applications. Moreover, there is no additional information about the parameter $\beta_g$.
\item The prior for the parameter $\gamma_g$ is a Jeffrey's distribution
that reflects the absence of knowledge about this parameter \cite{Jeffreys1946}. This
kind of prior is often used for scale parameters.
\end{itemize}
The resulting posterior distribution is therefore given by
\begin{equation}
\label{eq:post}
f(\vect{x},\vect{\theta}|\vect{y}) \propto \mathsf{1}_{C_{\delta}}(\vect{x}) \prod_{g=1}^G \left[ \left(
\frac{\beta_g}{2 \gamma_g^{1/\beta_g} \Gamma(1/\beta_g)}
\right)^{n_g}\exp \left( - \frac{1}{\gamma_g} \sum_{k\in S_g} |x_k|^{\beta_g}
  \right) \frac{1}{\gamma_g}
\mathsf{1}_{\mathbb{R}^+}(\gamma_g)\mathsf{1}_{[0,3]}(\beta_g)
\right].
\end{equation}

The Bayesian estimators (e.g., the maximum a posteriori (MAP) or
minimum mean square error (MMSE) estimators) associated with the posterior distribution \eqref{eq:post} have no simple
closed-form expression. The next section studies different sampling
strategies for generating samples distributed according
to the posterior distribution \eqref{eq:post}. 
 The generated samples will be used to estimate the unknown model parameter
and hyper-parameter vectors $\vect{x}$ and
$\vect{\theta}$.

\section{Sampling strategies} \label{sec:sampling}
This section proposes different MCMC methods to generate samples
distributed according to the posterior
$f(\vect{x},\vect{\theta}|\vect{y})$ defined in \eqref{eq:post}.

\subsection{Hybrid Gibbs Sampler}\label{sec:sampler1}
A very standard strategy to sample according to \eqref{eq:posterior}
is provided by the Gibbs sampler. The Gibbs sampler iteratively
generates samples distributed according to conditional
distributions associated with the target distribution. More precisely,
the basic Gibbs sampler iteratively generates samples distributed
according to $f(\vect{x}|\vect{\theta},\vect{y})$ and
$f(\vect{\theta}|\vect{x},\vect{y})$.
\subsubsection{Sampling the frame coefficients} \label{se:sampcoef1}\hfill \\
Straightforward calculations yield the following
conditional distribution
\begin{equation}
\label{eq:posterior_coef}
f(\vect{x}|\vect{\theta},\vect{y}) \propto \mathsf{1}_{C_{\delta}}(\vect{x})
\prod_{g=1}^G \exp \left( - \frac{1}{\gamma_g} \sum_{k\in S_g}
|x_k|^{\beta_g}
  \right)
\end{equation}
where $C_{\delta}$ is defined in \eqref{eq:Cdelta}.
This conditional distribution is a product of GG
distributions truncated on $C_{\delta}$. Actually, sampling
according to this truncated distribution is not always easy to
perform since the adjoint frame operator $F^*$ is
usually of large dimension. However, two alternative sampling strategies are detailed in what follows.
\paragraph{Naive sampling}\hfill \\
This sampling method proceeds by sampling according to independent
GG distributions
\begin{equation}\label{eq:naive}
\prod_{g=1}^G \exp \left( - \frac{1}{\gamma_g} \sum_{k\in S_g}
|x_k|^{\beta_g}
  \right)
\end{equation}
and then accepting the proposed candidate $\vect{x}$ only if
$N(\vect{y}-F^*\vect{x}) \leq \delta$. This method can be used for
any frame decomposition and any norm. However, it
can be quite inefficient because of a very low acceptance ratio,
especially when $\delta$ takes small values.

\paragraph{Gibbs sampler}\hfill \\
This sampling method is designed to sample more efficiently from the conditional distribution in
\eqref{eq:posterior_coef} when the considered frame is the union of $M$ orthonormal bases and
$N(.)$ is the Euclidean norm. In this case, the analysis frame operator
and the corresponding adjoint can be written as $F =
\left[\begin{array}{c}
F_1\\
\vdots\\
F_M\\
\end{array}\right]$
and $F^* = [F^*_1 \hdots F^*_M]$, respectively, where $\forall m
\in\{1, \hdots, M\}$, $F_m$ is the decomposition operator onto the
$m$th orthonormal basis such as
$F^*_m F_m = F_m F^*_m = \mathrm{I}$.\\
Every $\vect{x}\in \mathbb{R}^{K}$ with $K = ML$, can be decomposed as
$\vect{x} = \left[\vect{x}_1^{\trans},\ldots,\vect{x}_M^{\trans}\right]^{\trans}$
where $\forall m \in \{1, \ldots, M\},$ \linebreak 
$\vect{x}_m \in~\mathbb{R}^{L}$.\\

The Gibbs sampler for the generation of frame
coefficients draws vectors according to the conditional distribution
$f(\vect{x}_n|\vect{x}_{-n},\vect{y},\vect{\theta})$ under the
constraint $N(\vect{y}-F^*\vect{x})\leq\delta$, where
$\vect{x}_{-n}$ is the reduced size vector of dimension $\RR^{K-L}$
built from  $\vect{x}$ by removing the $n$th vector $\vect{x}_n$. If
$N(.)$ is the Euclidean norm, we have $\forall n \in \{1, \ldots, M\}$,
\begin{align}
&N(\vect{y} - \sum_{m=1}^M F^*_m \vect{x}_m) \leq \delta  \nonumber\\
\Leftrightarrow & \parallel F^*_n (F_n \vect{y} - \sum_{m=1}^M F_nF^*_m\vect{x}_m) \parallel \leq \delta\nonumber\\
\Leftrightarrow & \parallel F_n \vect{y} - \sum_{m \neq n} F_nF^*_m \vect{x}_m -\vect{x}_n \parallel \leq \delta \;\;\; (\mathrm{since}\;\; \forall \vect{z} \in \mathbb{R}^{L},\;\; \parallel F^*_n\vect{z} \parallel=\parallel\vect{z}\parallel)\nonumber\\
\Leftrightarrow & N(\vect{x}_n - \vect{c}_n) \leq \delta,
\end{align}
where $$\vect{c}_n = F_n \Big(\vect{y} - \sum_{m \neq n} F^*_m
\vect{x}_m\Big)\label{eq:cn}.$$ Having $\vect{x}_{-n}=(\vect{x}_m)_{m\neq n}$, it is thus easy
to compute the vector $\vect{c}_n$. To sample each
$\vect{x}_n$, we propose to use an MH step whose proposal distribution is supported on the
ball $B_{\vect{c}_n,\delta}$ defined by
\begin{equation}\label{eq:Bdelta}
B_{\vect{c}_n,\delta}=\{\vect{a}\in \mathbb{R}^L \mid N(\vect{a} - \vect{c}_n ) \leq \delta \}.
\end{equation}
Random generation from a pdf $q_{\delta}$ defined on
$B_{\vect{0},\delta}$ which has a simple expression is described in
Appendix~\ref{append:a1}. Having a closed form expression of this
pdf is important to be able to calculate the acceptance ratio of the
MH move. To take into account the value of $\vect{x}_n^{(i-1)}$
obtained at the previous iteration $(i-1)$, it may however be
preferable to choose a proposal distribution supported on a
restricted ball of radius $\eta \in ]0,\delta[$ containing
$\vect{x}_n^{(i-1)}$. This strategy similar to the
random walk MH algorithm \cite[p.~287]{Robert_04} 
results in a better exploration of regions associated with large
values of the conditional distribution
$f(\vect{x}|\vect{\theta},\vect{y})$.

More
precisely, we propose to choose a proposal distribution defined on
$B_{\hat{\vect{x}}_n^{(i-1)},\eta}$, where $\hat{\vect{x}}_n^{(i-1)}
= P(\vect{x}_n^{(i-1)}-\vect{c}_n)+ \vect{c}_n$ and $P$ is the
projection onto the ball $ B_{\vect{0},\delta-\eta}$ defined as

\begin{equation}
\forall \vect{a} \in \RR^L,\qquad
P(\vect{a}) =
\begin{cases}
\vect{a} & \mbox{if $N(\vect{a}) \le \delta-\eta$}\\
\displaystyle \frac{\delta-\eta}{N(\vect{a})}\vect{a} & \mbox{otherwise.}
\end{cases}
\end{equation}
This choice of the center of the ball guarantees that
$B_{\hat{\vect{x}}_n^{(i-1)},\eta}\subset B_{\vect{c}_n,\delta}$.
Moreover, any point of $B_{\vect{c}_n,\delta}$ can
be reached after consecutive draws in
$B_{\hat{\vect{x}}_n^{(i-1)},\eta}$. Note that the radius $\eta$ has
to be adjusted to ensure a good exploration of $B_{\vect{c}_n,\delta}$. In practice, it may also be interesting to fix a small enough value of $\eta$ so as to improve the acceptance ratio.

\noindent\underline{\textbf{Remark:}}\\
Alternatively, a Gibbs sampler can be used to draw successively the
$L$ elements $(x_{n,l})_{1 \le l \le L}$ of $\vect{x}_{n}$ under the
following constraint
\begin{align}
 & \parallel \vect{x}_n - \vect{c}_n \parallel \leq \delta\nonumber\\
\Leftrightarrow & -\sqrt{\delta^2-\sum_{k \neq l}(x_{n,k}-c_{n,k})^2} \leq x_{n,l}-c_{n,l} \leq \sqrt{\delta^2-\sum_{k \neq l}(x_{n,k}-c_{n,k})^2} ,\quad \forall l \in \{1,\ldots, L \} \nonumber\\
\end{align}
where $c_{n,k}$ is the $k$th element of the vector $\vect{c}_n$ (see \cite[p.133]{Liu_01} for related strategies).
However, this method is very time-consuming since
it proceeds sequentially for each component of the high dimensional
vector $\vect{x}$.%

\subsubsection{Sampling the hyper-parameter vector}\hfill \\
\label{sec:sampling_hyper} Instead of sampling $\vect{\theta}$
according to $f(\vect{\theta}|\vect{x},\vect{y})$, we propose to
iteratively sample according to
$f(\gamma_g|\beta_g,\vect{x},\vect{y})$ and
$f(\beta_g|\gamma_g,\vect{x},\vect{y})$. Straightforward
calculations allow us to obtain the following results
\begin{align}
f(\gamma_g|\beta_g,\vect{x},\vect{y})& \propto \gamma_g^{-\frac{n_g}{\beta_g}-1} \exp \left( - \frac{1}{\gamma_g} \sum_{k\in S_g}
|x_k|^{\beta_g} \right) \mathsf{1}_{\mathbb{R}^+}(\gamma_g), \\
f(\beta_g|\gamma_g,\vect{x},\vect{y}) & \propto
\frac{\beta_g^{n_g}}{\gamma_g^{n_g/\beta_g} \left[ \Gamma \left( 1/
\beta_g \right) \right]^{n_g}} \exp \left( - \frac{1}{\gamma_g}
\sum_{k\in S_g} |x_k|^{\beta_g} \right) \mathsf{1}_{[0,3]}(\beta_g).
\end{align}
Consequently, due to the new parameterization introduced in
\eqref{eq:priorreparam}, $f(\gamma_g|\beta_g,\vect{x},\vect{y})$ is
the pdf of the inverse gamma distribution $\mathcal{IG} \left(
\frac{n_g}{\beta_g}, \sum_{k\in S_g} |x_k|^{\beta_g}\right)$ that is
easy to sample. Conversely, it is more difficult to sample according
to the truncated pdf $f(\beta_g|\gamma_g,\vect{x},\vect{y})$. This
is achieved by using an MH move whose proposal
$q(\beta_g\mid\beta_g^{(i-1)})$ is a Gaussian distribution truncated
on the interval $[0,3]$ with standard deviation
$\sigma_{\beta_g}=0.05$ \cite{Dobigeon_TechReport_2007a}.
Note that the mode of this distribution is the
value of the parameter $\beta_g^{(i-1)}$ at the previous iteration
$(i-1)$.

The resulting method is the hybrid Gibbs
sampler summarized in Algorithm $1$.

\begin{algorithme}\label{algo:gibbs}
\vspace{-0.4cm}
\begin{itemize}
\item[\Pisymbol{pzd}{192}] Initialize with some $\vect{\theta}^{(0)}=(\vect{\theta}_g^{(0)})_{1\leq g\leq G}=(\gamma_g^{(0)},\beta_g^{(0)})_{1\leq g\leq G}$ and $\vect{x}^{(0)}\in C_\delta$, and set $i = 1$.
\item[\Pisymbol{pzd}{193}] Sampling $\vect{x}$ \\
For $n=1$ to $M$
\begin{itemize}
\item Compute $\vect{c}_n^{(i)} = F_n \Big(\vect{y} - \sum_{m<n} F^*_m \vect{x}_m^{(i)}-\sum_{m>n} F^*_m \vect{x}_m^{(i-1)}\Big)$\\ and
 $\hat{\vect{x}}_n^{(i-1)} = P(\vect{x}_n^{(i-1)}-\vect{c}_n^{(i)})
+\vect{c}_n^{(i)}$.
\item  Simulate $\vect{x}_n^{(i)}$ as follows:
\begin{itemize}
\item Generate $\widetilde{\vect{x}}_n^{(i)} \sim
q_\eta(\vect{x}_n-\hat{\vect{x}}_n^{(i-1)})$ where $q_\eta$ is
defined on $B_{\vect{0},\eta}$  (see Appendix \ref{append:a1}).
\item Compute the ratio $$\textstyle r(\widetilde{\vect{x}}_n^{(i)},\vect{x}_n^{(i-1)})
= \frac{f(\widetilde{\vect{x}}_n^{(i)}|\vect{\theta}^{(i-1)},(\vect{x}_m^{(i)})_{m<n},(\vect{x}_m^{(i-1)})_{m>n},\vect{y})\; q_\eta\big(\vect{x}_n^{(i-1)}-
P(\widetilde{\vect{x}}_n^{(i)}-\vect{c}_n^{(i)})-\vect{c}_n^{(i)}\big)}{f(\vect{x}_n^{(i-1)}|\vect{\theta}^{(i-1)},(\vect{x}_m^{(i)})_{m<n},(\vect{x}_m^{(i-1)})_{m>n},\vect{y})\; q_\eta\big(\widetilde{\vect{x}}_n^{(i)}-\hat{\vect{x}}_n^{(i-1)}\big)}$$
and accept the proposed candidate with the probability $\min\{1,r(\widetilde{\vect{x}}_n^{(i)},\vect{x}_n^{(i-1)})\}$.
\end{itemize}
\end{itemize}
\item[\Pisymbol{pzd}{194}] Sampling $\vect{\theta}$\\
For $g=1$ to $G$
\begin{itemize}
\item Generate $\gamma_g^{(i)} \backsim \mathcal{IG} \left(
\frac{n_g}{\beta_g^{(i-1)}}, \sum_{k\in S_g} |x_k^{(i)}|^{\beta_g^{(i-1)}}\right)$.
\item Simulate $\beta_g^{(i)}$
as follows:
\begin{itemize}
\item Generate $\widetilde{\beta}_g^{(i)} \backsim q(\beta_g\mid \beta_g^{(i-1)})$
\item Compute the ratio $$r(\widetilde{\beta}_g^{(i)},\beta_g^{(i-1)})=\frac{f(\widetilde{\beta}_g^{(i)}|\gamma_g^{(i)},\vect{x}^{(i)},\vect{y})q(\beta_g^{(i-1)}\mid\widetilde{\beta}_g^{(i)})}{f(\beta_g^{(i-1)}|\gamma_g^{(i)},\vect{x}^{(i)},\vect{y})q(\widetilde{\beta}_g^{(i)}\mid \beta_g^{(i-1)})}$$ and accept the proposed candidate with the probability $\min\{1,r(\widetilde{\beta}_g^{(i)},\beta_g^{(i-1)})\}$.
\end{itemize}

\end{itemize}
\item[\Pisymbol{pzd}{195}] Set $i \leftarrow i+1$ and goto \Pisymbol{pzd}{193} until convergence.
\end{itemize}
\caption{Proposed Hybrid Gibbs sampler to simulate according to
$f(\vect{x},\vect{\theta}|\vect{y})$ (superscript $\cdot^{(i)}$ indicates values computed at iteration number
$i$).}
\end{algorithme}
Although this algorithm is intuitive and simple to implement, it
must be pointed out that it was derived under the restrictive
assumption that the considered frame is the union of $M$ orthonormal bases.
When this assumption does not hold, another algorithm proposed
in the next section allows us to sample frame coefficients and the
related hyper-parameters by exploiting algebraic properties of
frames.

\subsection{Hybrid MH sampler using algebraic properties of frame representations}\label{sec:sampler2}
As a direct generation of samples according to
$f(\vect{x}|\vect{\theta},\vect{y})$ is generally impossible, we
propose here an alternative that replaces the Gibbs move by
an MH move. This MH move aims at
sampling globally a candidate $\vect{x}$ according to a proposal
distribution. This candidate is accepted or rejected with the
standard MH acceptance ratio. The efficiency of the MH move strongly
depends on the choice of the proposal distribution for $\vect{x}$.
We denote as $\vect{x}^{(i)}$ the $i$th accepted sample of the
algorithm and $q(\vect{x}\mid \vect{x}^{(i-1)})$ the proposal that
is used to generate a candidate at iteration $i$. The main
difficulty for choosing $q(\vect{x}\mid\vect{x}^{(i-1)})$ stems from
the fact that it must guarantee that $\vect{x} \in C_\delta$ (as
mentioned in Section \ref{sec:FD}) while yielding a tractable
expression of $q(\vect{x}^{(i-1)} \mid \vect{x})/q(\vect{x} \mid
\vect{x}^{(i-1)})$.

For this reason, we propose to exploit the algebraic properties of frame
representations. More precisely, any frame coefficient vector can be
decomposed as $\vect{x}=\vect{x}_H+\vect{x}_{H^\perp}$, where
$\vect{x}_{H}$ and $\vect{x}_{H^\perp}$ are realizations of random vectors taking
their values in $H=\mathrm{Ran}(F)$ and
$H^{\perp}=[\mathrm{Ran}(F)]^{\perp}=\mathrm{Null}(F^*)$,
respectively.\footnote{We recall that the range of $F$ is
$\mathrm{Ran}(F) = \{ \vect{x}\in \mathbb{R}^K | \exists \vect{y}
 \in \RR^L, F\vect{y} = \vect{x}\}$
and the null space of $F^*$ is $\mathrm{Null}(F^*) = \{\vect{x} \in
\mathbb{R}^K| F^*\vect{x} = \vect{0}\}$.} The proposal distribution
used in this paper allows us to generate samples $\vect{x}_{H} \in
H$ and $\vect{x}_{H^{\perp}} \in H^{\perp}$. More precisely, the
following separable form of the proposal pdf will be considered
\begin{equation}
q(\vect{x}\mid \vect{x}^{(i)})=q \left(\vect{x}_{H}\mid
\vect{x}_H^{(i-1)} \right)\,q \left(\vect{x}_{H^{\perp}}\mid
\vect{x}_{H^{\perp}}^{(i-1)} \right)
\end{equation}
where $\vect{x}_H^{(i-1)} \in H$, $\vect{x}_{H^{\perp}}^{(i-1)} \in
H^{\perp}$ and
$\vect{x}^{(i-1)}=\vect{x}_H^{(i-1)}+\vect{x}_{H^{\perp}}^{(i-1)}$. In
other words, independent sampling of $\vect{x}_{H}$
and $\vect{x}_{H^{\perp}}$ will be performed.

If we consider the decomposition
$\vect{x}=\vect{x}_{H}+\vect{x}_{H^{\perp}}$, sampling $\vect{x}$
in  $C_{\delta}$ is equivalent to sampling $\vect{\lambda} \in
\overline{C}_{\delta}$, where
$\overline{C}_{\delta}=\{\vect{\lambda}\in \mathbb{R}^L|
N(\vect{y}-F^*F\vect{\lambda})\leq \delta\}$ 
 is the inverse image of $C_{\delta}$ under $F$.\\
Indeed, we can write
$\vect{x}_{H}=F\vect{\lambda}$ where $\vect{\lambda}\in
\mathbb{R}^L$ and,  since $\vect{x}_{H^{\perp}}\in
\mathrm{Null}(F^*)$, $F^*\vect{x}=F^*F\vect{\lambda}$. Sampling
$\vect{\lambda}$ in  $\overline{C}_{\delta}$ can be easily achieved, e.g., by
generating $\vect{u}$ from a distribution supported on the ball
$B_{\vect{y},\delta}$
and by taking $\vect{\lambda}=(F^*F)^{-1}\vect{u}$. \\
To make the sampling of $\vect{x}_{H}$ at iteration $i$ more efficient, taking into account the sampled value at the previous iteration $\vect{x}^{(i-1)}_{H} = F\vect{\lambda}^{(i-1)}= F(F^*F)^{-1}\vect{u}^{(i-1)}$ may be interesting.
Similarly to Section~\ref{se:sampcoef1}b), and to random walk generation techniques, we proceed by generating randomly $\vect{u}$ in $B_{\hat{\vect{u}}^{(i-1)},\eta}$ where $\eta \in ]0,\delta[$ and $\hat{\vect{u}}^{(i-1)} = P(\vect{u}^{(i-1)}-\vect{y})+\vect{y}$. This allows us
to draw a vector $\vect{u}$ such that $\vect{x}_H = F(F^*F)^{-1}\vect{u} \in C_\delta$ and $N(\vect{u}- \vect{u}^{(i-1)})\le 2 \eta$.  The generation of $\vect{u}$ can then be performed as explained in
Appendix~\ref{append:a1} provided that $N(.)$ is an $\ell^p$ norm with $p\in [1,+\infty]$.

Once we have simulated $\vect{x}_{H}=F\vect{\lambda} \in H\cap
C_\delta$ (which ensures that $\vect{x}$ is in $C_{\delta}$),
$\vect{x}_{H^{\perp}}$ has to be sampled as an element of
$H^{\perp}$. Since $\vect{y}=F^* \vect{x} +\vect{n}=F^* \vect{x}_{H}
+\vect{n}$, there is no information in $\vect{y}$ about
$\vect{x}_{H^{\perp}}$. As a consequence, and for simplicity reasons, we propose to sample
$\vect{x}_{H}$ by drawing $\vect{z}$ according to the Gaussian distribution
$\mathcal{N}(\vect{x}^{(i-1)},\sigma_\vect{x}^2{\boldsymbol I})$ and
by projecting $\vect{z}$ onto $H^{\perp}$, i.e.,
\begin{equation}
\vect{x}_{H^{\perp}}=\Pi_{H^{\perp}} \vect{z}
\end{equation}
where $\Pi_{H^{\perp}} = \mathrm{I}-F(F^*F)^{-1}F^*$ is the orthogonal
projection operator onto $H^{\perp}$.

Note here that using a tight frame makes the computation of both $\vect{x}_H$
and $\vect{x}_{H^{\perp}}$ much easier due to the relation $F^*F=\mu \mathrm{I}$.

Let us now derive the expression of the proposal pdf.
It can be noticed that, if $K > L$,
there exists a linear operator $F_\perp$ from $\RR^{K-L}$ to $\RR^L$ which is
semi-orthogonal (i.e., $F_\perp^* F_\perp = \mathrm{I}$) and
orthogonal to $F$ (i.e., $F_\perp^* F = 0$), such that
\begin{equation}
\vect{x} = \underbrace{\,F \vect{\lambda}\,}_{\vect{x}_H}\;+\;
\underbrace{\,F_\perp \vect{\lambda}_\perp\,}_{\vect{x}_{H^\perp}}
\end{equation}
and $\vect{\lambda}_\perp = F_\perp^* \vect{x} \in \RR^{K-L}$.
Standard rules on bijective linear transforms of random vectors lead to
\begin{equation}
q(\vect{x} \mid \vect{x}^{(i-1)}) =
| \operatorname{det}\big([F\quad F_\perp]\big)|^{-1}
q(\vect{\lambda} \mid \vect{x}^{(i-1)}) q(\vect{\lambda}_\perp \mid
\vect{x}^{(i-1)})
\end{equation}
where, due to the bijective linear mapping between $\vect{\lambda}$
and $\vect{u} = F^*F\vect{\lambda}$
\begin{equation}
q(\vect{\lambda} \mid \vect{x}^{(i-1)}) = \operatorname{det}(FF^*)\;
q_\eta(\vect{u}-\hat{\vect{u}}^{(i-1)})
\end{equation}
and $q(\vect{\lambda}_\perp \mid \vect{x}^{(i-1)})$ is the pdf of
the Gaussian distribution
$\mathcal{N}(\vect{\lambda}_\perp^{(i-1)},\sigma_\vect{x}^2{\boldsymbol
I})$ with mean $\vect{\lambda}_\perp^{(i-1)} = F_\perp^*
\vect{x}^{(i-1)}$. Recall that $q_\eta$ denotes a distribution on
the ball $B_{\vect{0},\eta}$ as expressed in Appendix
\ref{append:a1}. Due to the symmetry of the Gaussian distribution,
it can be deduced that
\begin{equation}
\frac{q(\vect{x}^{(i-1)}\mid\vect{x})}{q(\vect{x}\mid\vect{x}^{(i-1)
})}
= \frac{q_\eta(\vect{u}^{(i-1)}-P(\vect{u}-\vect{y})-\vect{y})}{q_\eta(\vect{u}-\hat{\vect{u}}^{(i-1)})}.
\end{equation}
This expression remains valid in the degenerate case when $K = L$
(yielding
$\vect{x}_{H^\perp} = \vect{0}$). Finally, it is important to note
that, if $q_\eta$ can be chosen as a uniform distribution on the
ball $B_{\vect{0},\eta}$, the above ratio reduces to
$1$, which simplifies the computation of the MH acceptance ratio.\\
 The final algorithm is summarized in Algorithm 2.
Note that the sampling of the hyper-parameter vector is performed as
for the hybrid Gibbs sampler in Section \ref{sec:sampling_hyper}.

\begin{algorithme} \label{algo:main}
\vspace{-0.4cm}
\begin{itemize}
\item[\Pisymbol{pzd}{192}] Initialize with some $\vect{\theta}^{(0)}=(\vect{\theta}_g^{(0)})_{1\leq g\leq G}=(\gamma_g^{(0)},\beta_g^{(0)})_{1\leq g\leq G}$ and $\vect{u}^{(0)} \in B_{\vect{y},\delta}$.
Set $\vect{x}^{(0)} = F(F^*F)^{-1} \vect{u}^{(0)}$ and $i = 1$.
\item[\Pisymbol{pzd}{193}] Sampling $\vect{x}$
\begin{itemize}
\item Compute $\hat{\vect{u}}^{(i-1)} = P(\vect{u}^{(i-1)}-\vect{y})+\vect{y}$.
\item Generate $\widetilde{\vect{u}}^{(i)} \backsim q_\eta(\vect{u}-\hat{\vect{u}}^{(i-1)})$ where $q_\eta$ is defined on $B_{\vect{0},\eta}$ (see Appendix \ref{append:a1}).
\item Compute $\widetilde{\vect{x}}_H^{(i)} = F(F^*F)^{-1}\widetilde{\vect{u}}^{(i)}$.
\item Generate $z^{(i)}\backsim \mathcal{N}(\vect{x}^{(i-1)},\sigma_\vect{x}^2{\boldsymbol I})$.
\item Compute $\widetilde{\vect{x}}_{H^\perp}^{(i)} = \Pi_{H^\perp} z^{(i)}$
and $\widetilde{\vect{x}}^{(i)} = \widetilde{\vect{x}}_H^{(i)}+\widetilde{\vect{x}}_{H^\perp}^{(i)}$.
\item Compute the ratio
\[
r(\widetilde{\vect{x}}^{(i)},\vect{x}^{(i-1)})= \dfrac{
f(\widetilde{\vect{x}}^{(i)}|\vect{\theta}^{(i-1)},\vect{y}) \;
q_\eta\big(\vect{u}^{(i-1)}-P(\widetilde{\vect{u}}^{(i)}-\vect{y})-\vect{y}\big)}
{f(\vect{x}^{(i-1)}|\vect{\theta}^{(i-1)},\vect{y})\;
q_\eta\big(\widetilde{\vect{u}}^{(i)}-\hat{\vect{u}}^{(i-1)}\big)}
\]
and accept the proposed candidates $\widetilde{\vect{u}}^{(i)}$ and
$\widetilde{\vect{x}}^{(i)}$
 with probability
$\min\{1,r(\widetilde{\vect{x}}^{(i)},\vect{x}^{(i-1)})\}$.
\end{itemize}
\item[\Pisymbol{pzd}{194}] Sampling $\vect{\theta}$\\
For $g=1$ to $G$
\begin{itemize}
\item Generate $\gamma_g^{(i)} \backsim \mathcal{IG} \left(
\frac{n_g}{\beta_g^{(i-1)}}, \sum_{k\in S_g} |x_k^{(i)}|^{\beta_g^{(i-1)}}\right)$.
\item Simulate $\beta_g^{(i)}$
as follows
\begin{itemize}
\item Generate $\widetilde{\beta}_g^{(i)} \backsim q(\beta_g\mid \beta_g^{(i-1)})$
\item Compute the ratio $$r(\widetilde{\beta}_g^{(i)},\beta_g^{(i-1)})=\frac{f(\widetilde{\beta}_g^{(i)}|\gamma_g^{(i)},\vect{x}^{(i)},\vect{y})q(\beta_g^{(i-1)}\mid\widetilde{\beta}_g^{(i)})}{f(\beta_g^{(i-1)}|\gamma_g^{(i)},\vect{x}^{(i)},\vect{y})q(\widetilde{\beta}_g^{(i)}\mid \beta_g^{(i-1)})}$$ and accept the proposed candidate with the probability $\min\{1,r(\widetilde{\beta}_g^{(i)},\beta_g^{(i-1)})\}$.
\end{itemize}
\end{itemize}
\item[\Pisymbol{pzd}{195}] Set $i \leftarrow i+1$ and goto \Pisymbol{pzd}{193} until convergence.
\end{itemize}
\caption{Proposed Hybrid MH sampler using algebraic properties of frame representations to simulate according to
$f(\vect{x},\vect{\theta}|\vect{y})$.}
\end{algorithme}

Experimental estimation results and applications to some image denoising problems
of the proposed stochastic sampling techniques are provided in the next section.
\section{Simulation Results}
\label{sec:simuls}
\subsection{Validation experiments}
\label{sec:validation} 
\subsubsection{Example 1}\hfill\\
\label{sec:Val_Ex1}
To show the effectiveness of our algorithm, a
first set of experiments was carried out on synthetic images. As a
frame representation, we used the union of two 2D separable wavelet
bases $\mathcal{B}_1$ and $\mathcal{B}_2$ using Daubechies and
shifted Daubechies filters of length 8 and 4, respectively. The
$\ell_2$ norm was used for $N(\cdot)$ in (\ref{eq:delta}) with
$\delta=10^{-4}$. To generate a synthetic image, we synthesized
wavelet frame coefficients $\vect{x}$ from known prior
distributions.

Let $\vect{x_1}=(a_1,(h_{1,j},v_{1,j},d_{1,j})_{1\leq j\leq 2})$ and
$\vect{x_2}=(a_2,(h_{2,j},v_{2,j},d_{2,j})_{1\leq j \leq 2})$ be the
sequences of wavelet basis coefficients generated in $\mathcal{B}_1$
and $\mathcal{B}_2$, where $a, h,v,d$ stand for approximation,
horizontal, vertical and diagonal coefficients and the index $j$ designates the
resolution level. Wavelet frame coefficients have been generated
from a GG distribution in accordance with the chosen priors. The
coefficients in each subband have been modeled with the same values
of the  hyper-parameters $\alpha_g$ and $\beta_g$, which means that
each subband forms a group of index $g$. The number of groups (i.e. the
number of subbands) $G$ is therefore equal to $14$. A uniform prior
distribution over $[0,3]$ has been chosen for parameter $\beta_g$
whereas a Jeffrey's prior has been assigned to each parameter
$\gamma_g$. 

After generating the hyper-parameters from their prior
distributions, a set of frame coefficients is randomly generated to
synthesize the observed data. The hyper-parameters
are then supposed unknown, sampled using the proposed algorithm, and
estimated by computing the mean of the generated samples according
to the MMSE principle. Having reference values, the normalized mean
square erors (NMSEs) related to the estimation of each
hyper-parameter belonging to a given group (here a given
subband) have been computed from $30$ Monte Carlo runs. The NMSEs
computed for the estimators associated with the two samplers of
Sections \ref{sec:sampler1} and \ref{sec:sampler2} are reported in
Table~\ref{tab:error}. 
\newpage
\begin{table}[!ht]
\centering \caption{NMSEs for the estimated hyper-parameters (30 runs).}
\begin{tabular}{|l|c|c|c|c|}
\hline
& \multicolumn{4}{|c|}{NMSE} \\
\cline{2-5}
& \multicolumn{2}{|c|}{Sampler 1}& \multicolumn{2}{|c|}{Sampler 2} \\
\cline{2-5}
& $\beta$ &  $\alpha$&  $\beta$& $\alpha$ \\
\hline
 $h_{1,1}$&0.015& 0.006&0.012& 0.030\\
\hline
 $v_{1,1}$&0.022 & 0.021 & 0.022& 0.026\\
\hline
 $d_{1,1}$&0.06& 0.016 & 0.011& 0.044\\
\hline
 $h_{1,2}$& 0.04& 0.003 & 0.021& 0.026\\
\hline
 $v_{1,2}$& 0.020& 0.027& 0.020& 0.019\\
\hline
 $d_{1,2}$& 0.013 & 0.016& 0.023& 0.041\\
\hline
 $a_1$& 0.039 &0.08 & 0.039& 0.023\\
\hline
\hline
 $h_{2,1}$&0.015 &0.030 & 0.015& 0.025\\
\hline
 $v_{2,1}$& 0.051 & 0.07& 0.025& 0.031\\
\hline
 $d_{2,1}$& 0.027 & 0.039& 0.029& 0.023\\
\hline
 $h_{2,2}$& 0.040 &0.024 & 0.016& 0.034\\
\hline
 $v_{2,2}$& 0.08 & 0.019& 0.013& 0.022\\
\hline
 $d_{2,2}$& 0.05 & 0.015& 0.011& 0.040\\
\hline
 $a_2$&0.010 & 0.064& 0.010& 0.028\\
\hline
\end{tabular}
\label{tab:error}
\end{table}

Table~\ref{tab:error} shows that the proposed
algorithms (using Sampler $1$ of Section \ref{sec:sampler1} and
Sampler $2$ of Section \ref{sec:sampler2}) provide accurate
estimates of the hyper-parameters. The two samplers perform
similarly for this experiment. However, one advantage of Sampler 2 is that it
can be
applied to different kinds of redundant frames, unlike Sampler 1. Indeed, as
reported in Section \ref{sec:sampler1}, the conditional distribution
\eqref{eq:posterior_coef} is generally difficult to sample when the frame
representation is not the union of orthonormal bases. 

\begin{figure}[!ht]
\centering
\begin{tabular}{c c}
$a_1$: $\beta = 1.7$, $\gamma = 104$ & $h_{1,2}$: $\beta =1.98 $, $\gamma =143.88 $\\
\includegraphics[width=6cm, height=4cm]{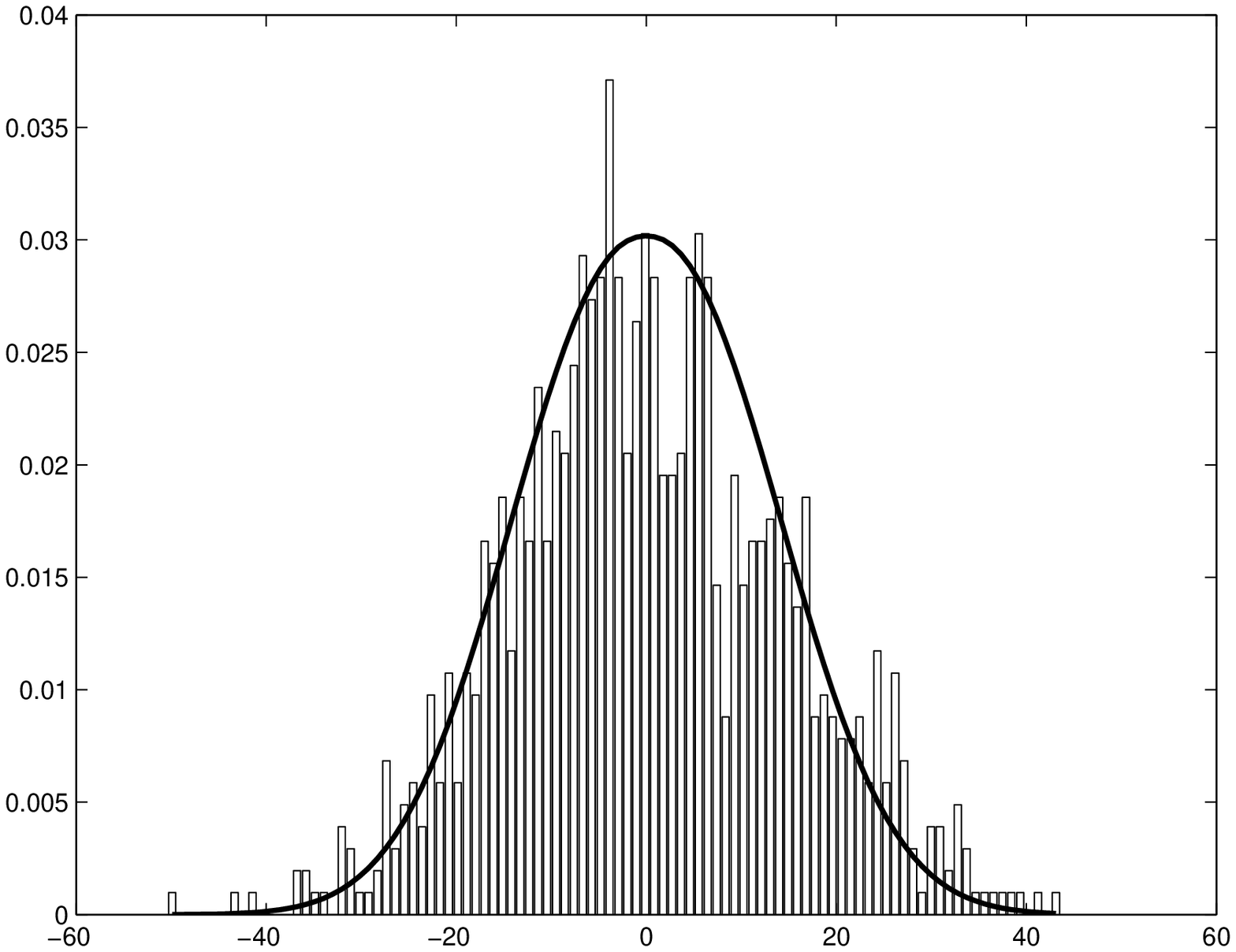} &
\includegraphics[width=6cm, height=4cm]{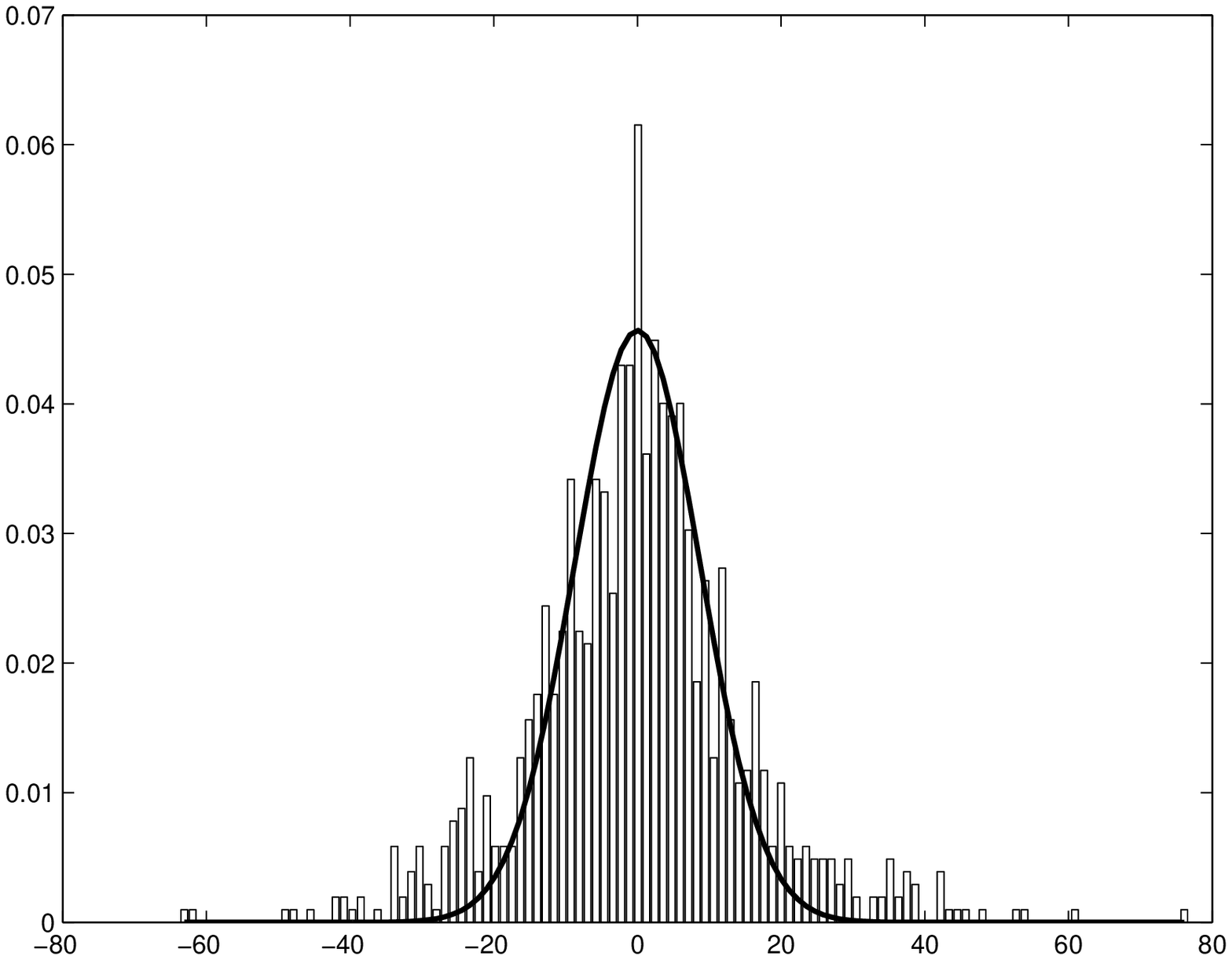}
\end{tabular}
\vspace{-0.2cm}
\caption{Examples of empirical approximation (left) and detail (right) histograms and pdfs of frame coefficients corresponding to a synthetic image.}
\label{fig:histograms1}
\end{figure}

Two examples of empirical histograms of known reference wavelet
frame coefficients (corresponding to $\mathcal{B}_1$) and pdfs with estimated hyper-parameters are
plotted in Fig.~\ref{fig:histograms1} to illustrate the good
performance of the estimator. 
\newpage

\subsubsection{Example 2}\hfill \\
In this experiment, another frame representation is considered, namely a tight frame version of the translation invariant wavelet transform \cite{Coifman_R_1995_was_tra_id} with Daubechies filters of length 8. The $\ell_2$ norm was also used for $N(.)$ in (\ref{eq:delta}) with
$\delta=10^{-4}$. Let $\vect{x}=(a,(h_{j},v_{j},d_{j})_{1\leq j\leq 2})$ denote the frame coefficients vector. We used the same process to generate frame coefficients as for Example 1. The coefficients in each subband (i.e. each group) have been modeled with the same values of the  hyper-parameters $\gamma_g$ and $\beta_g$, the number of groups being equal to $7$. The same priors for the hyper-parameters $\gamma_g$ and $\beta_g$ as for Example 1 have been used.\\
After generating the hyper-parameters and frame coefficients, the hyper-parameters
are then supposed unknown, sampled using the proposed algorithm, and
estimated using the MMSE estimator. 
Table \ref{tab:errorond} shows NMSEs based on reference values of each hyper-parameter. Note that Sampler 1 is difficult to be implemented in this case because of the used frame properties. Consequently, only NMSE values for Sampler 2 have been reported in Table \ref{tab:errorond}.

\begin{table}[!ht]
\centering \caption{NMSEs for the estimated hyper-parameters using Sampler 2 (30 runs).}
\begin{tabular}{|l|c|c|}
\hline
& \multicolumn{2}{|c|}{NMSE} \\
\cline{2-3}
\cline{2-3}
& $\beta$ &  $\alpha$\\
\hline
 $h_{1}$&0.05& 0.027\\
\hline
 $v_{1}$&0.024 & 0.007\\
\hline
 $d_{1}$&0.05& 0.014 \\
\hline
 $h_{2}$& 0.037& 0.028\\
\hline
 $v_{2}$& 0.051& 0.044\\
\hline
 $d_{2}$& 0.04 & 0.012\\
\hline
 $a$& 0.04 &0.05\\
\hline
\end{tabular}
\label{tab:errorond}
\end{table}

\subsection{Convergence results}
\begin{figure}[!ht]
\centering
\begin{tabular}{c c}
\centering
Chain $1$ & Chain $2$\\
\includegraphics[width=8cm, height=3.4cm]{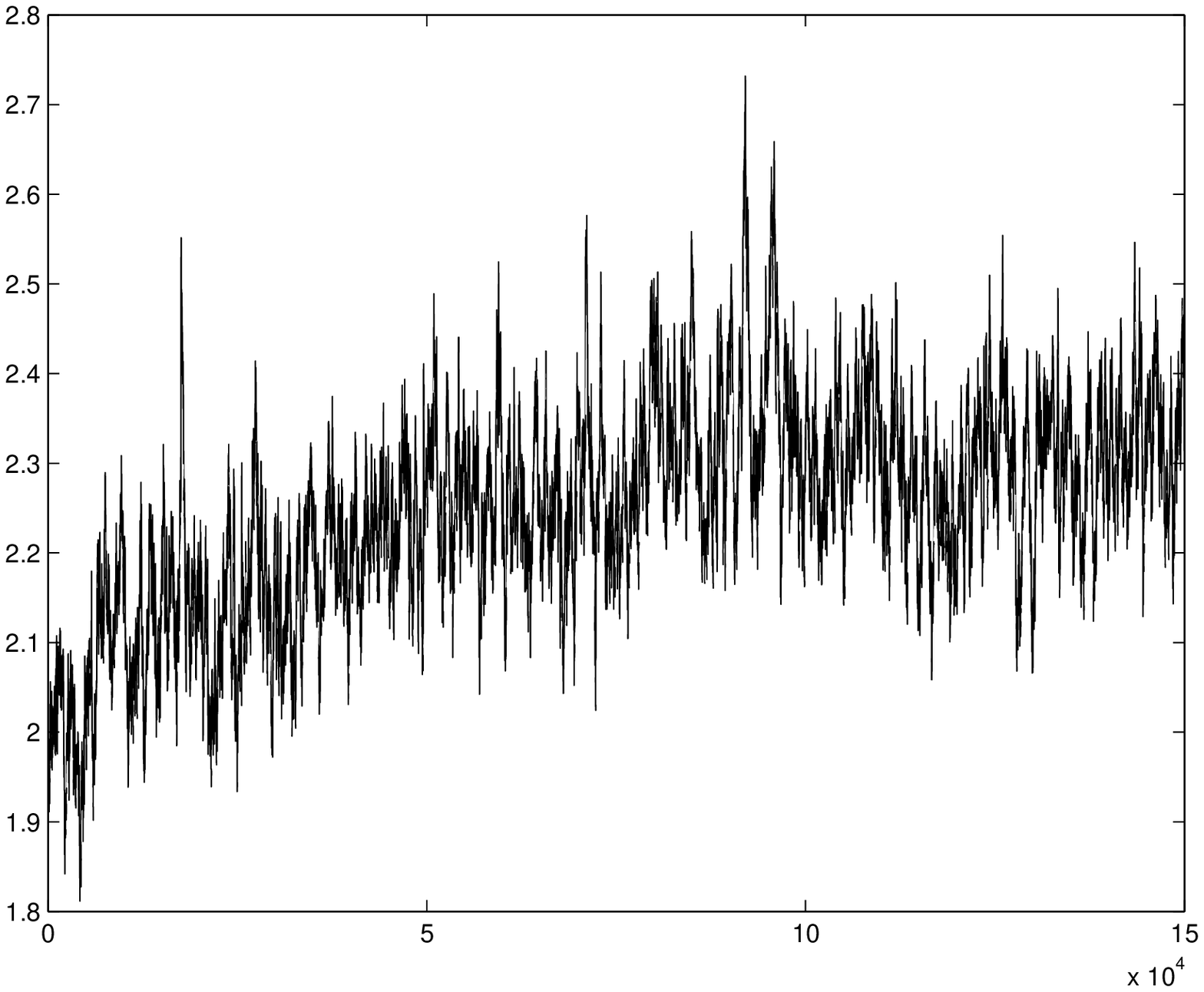} &
\includegraphics[width=8cm, height=3.4cm]{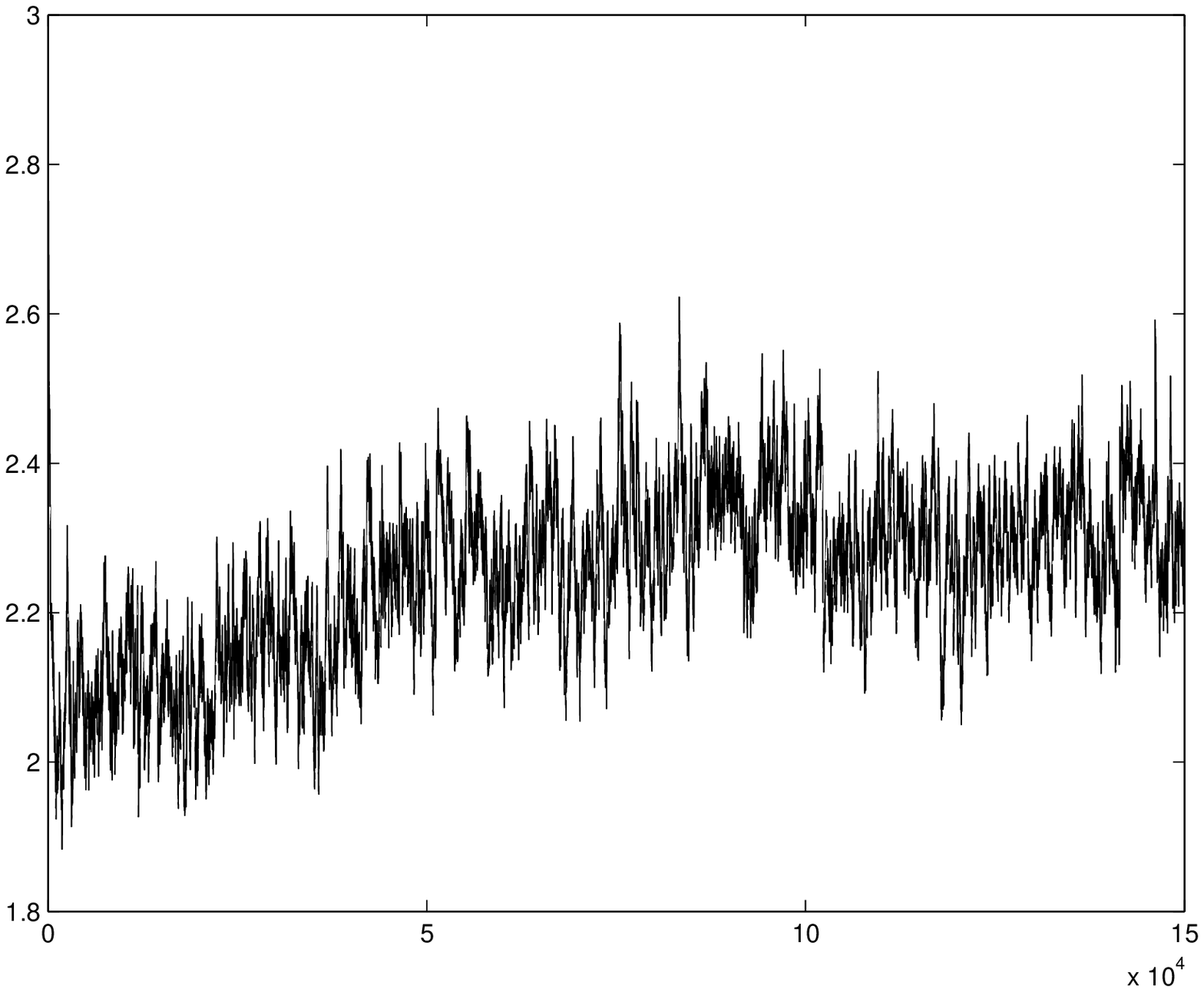}\\
\multicolumn{2}{c}{$\beta = 2,3$ - $\mathrm{PSRF} = 1.02$} \\
\includegraphics[width=8cm, height=3.4cm]{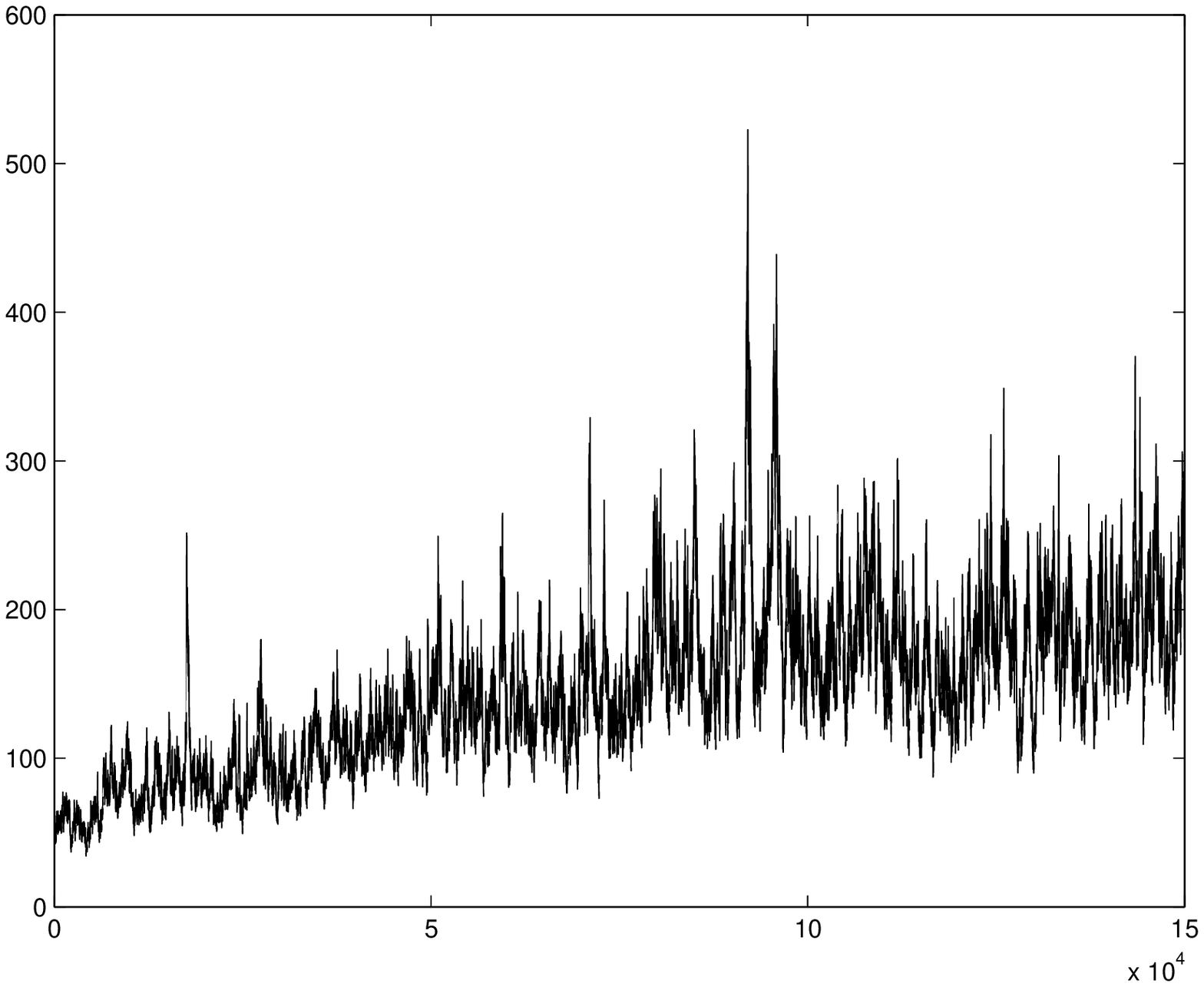}&
\includegraphics[width=8cm, height=3.4cm]{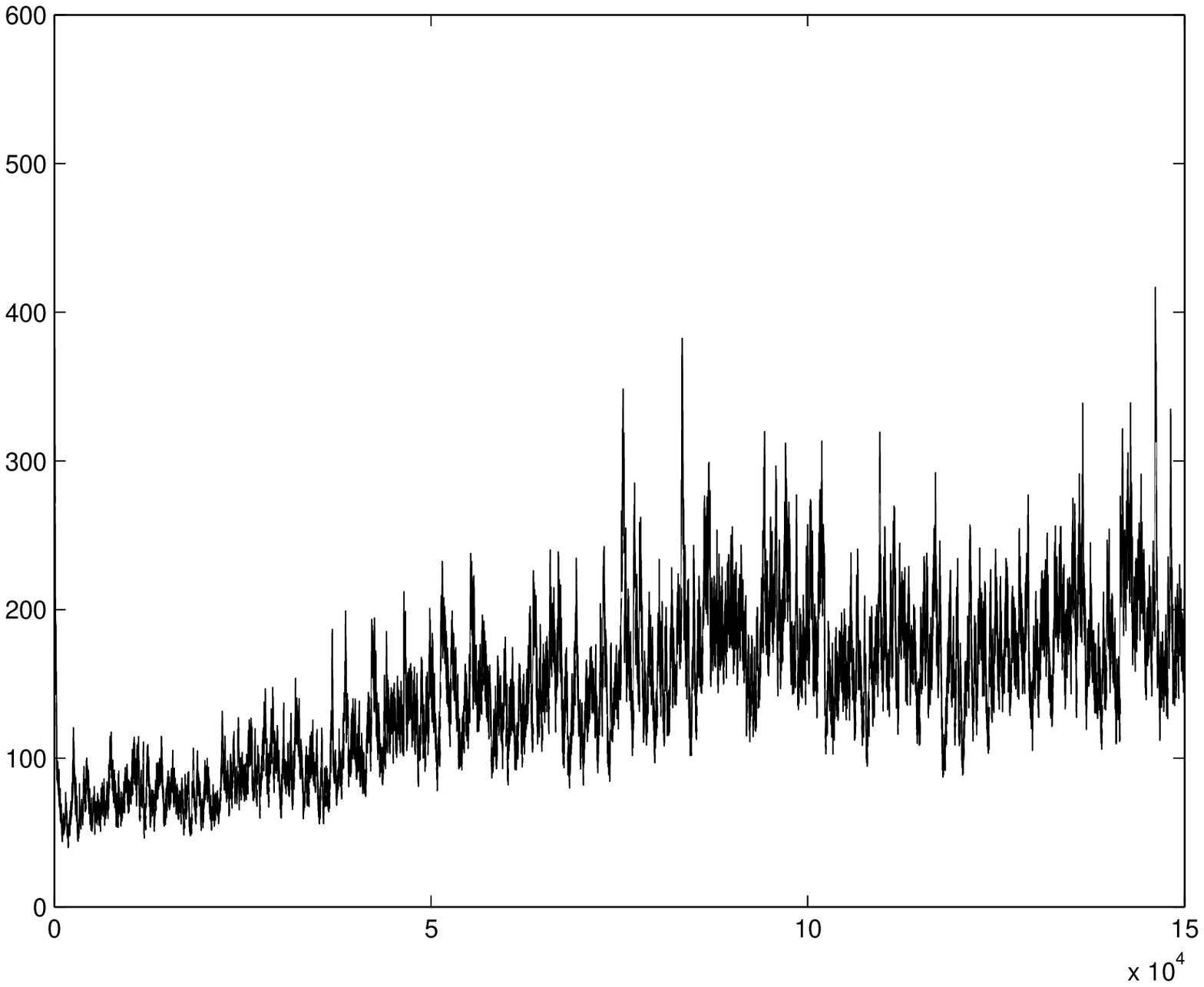} \\
\multicolumn{2}{c}{$\gamma=185$ - $\mathrm{PSRF} = 0.96$} \\
\end{tabular}
\caption{Ground truth values and sample path for the hyper-parameters $\beta$ and $\gamma$
related to $v_{1,1}$ in $\mathcal{B}_1$.} \label{fig:cv1}
\end{figure}

\begin{figure}[!ht]
\centering
\begin{tabular}{c c}
\centering
Chain $1$ & Chain $2$\\
\includegraphics[width=8cm, height=3.4cm]{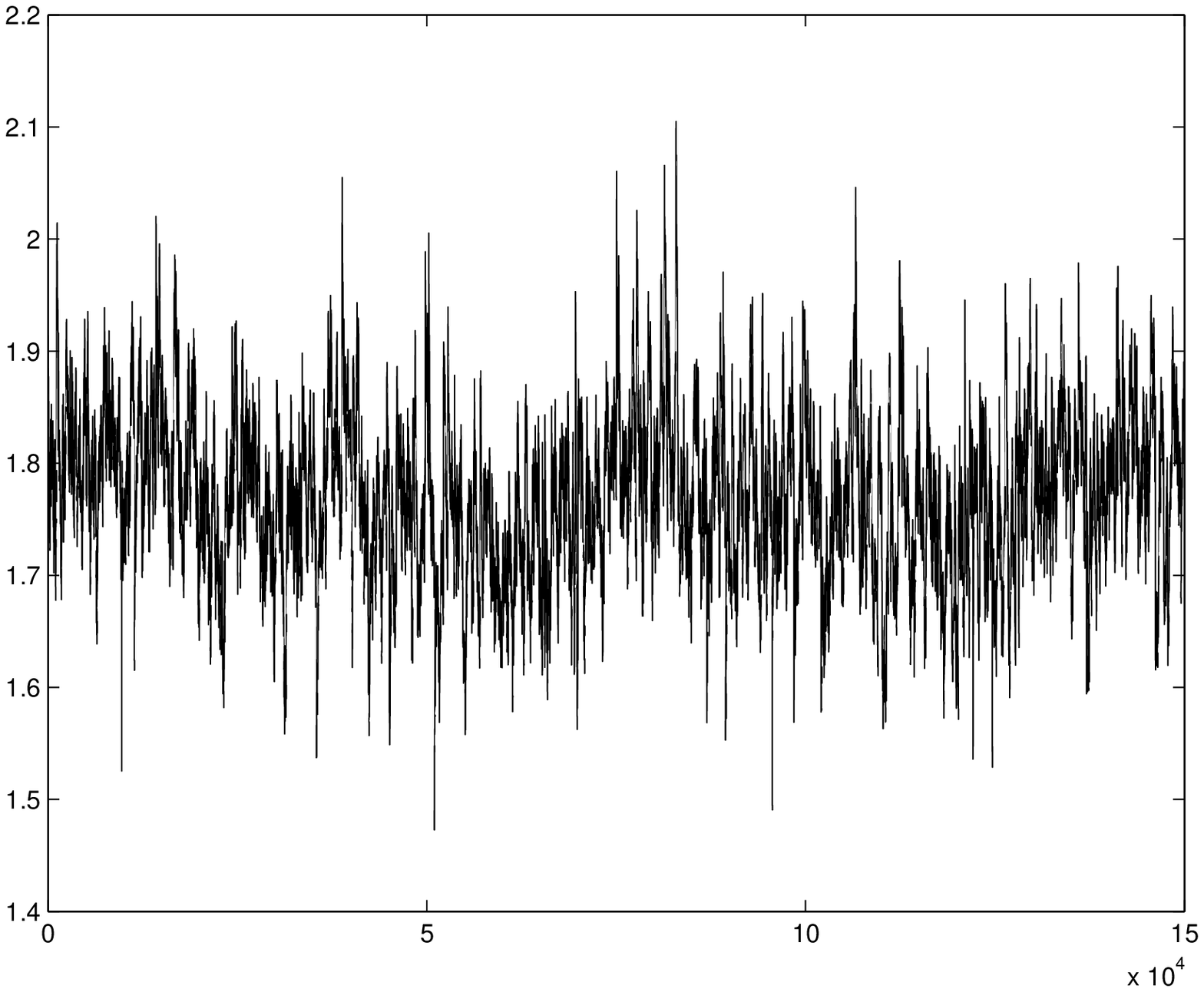} &
\includegraphics[width=8cm, height=3.4cm]{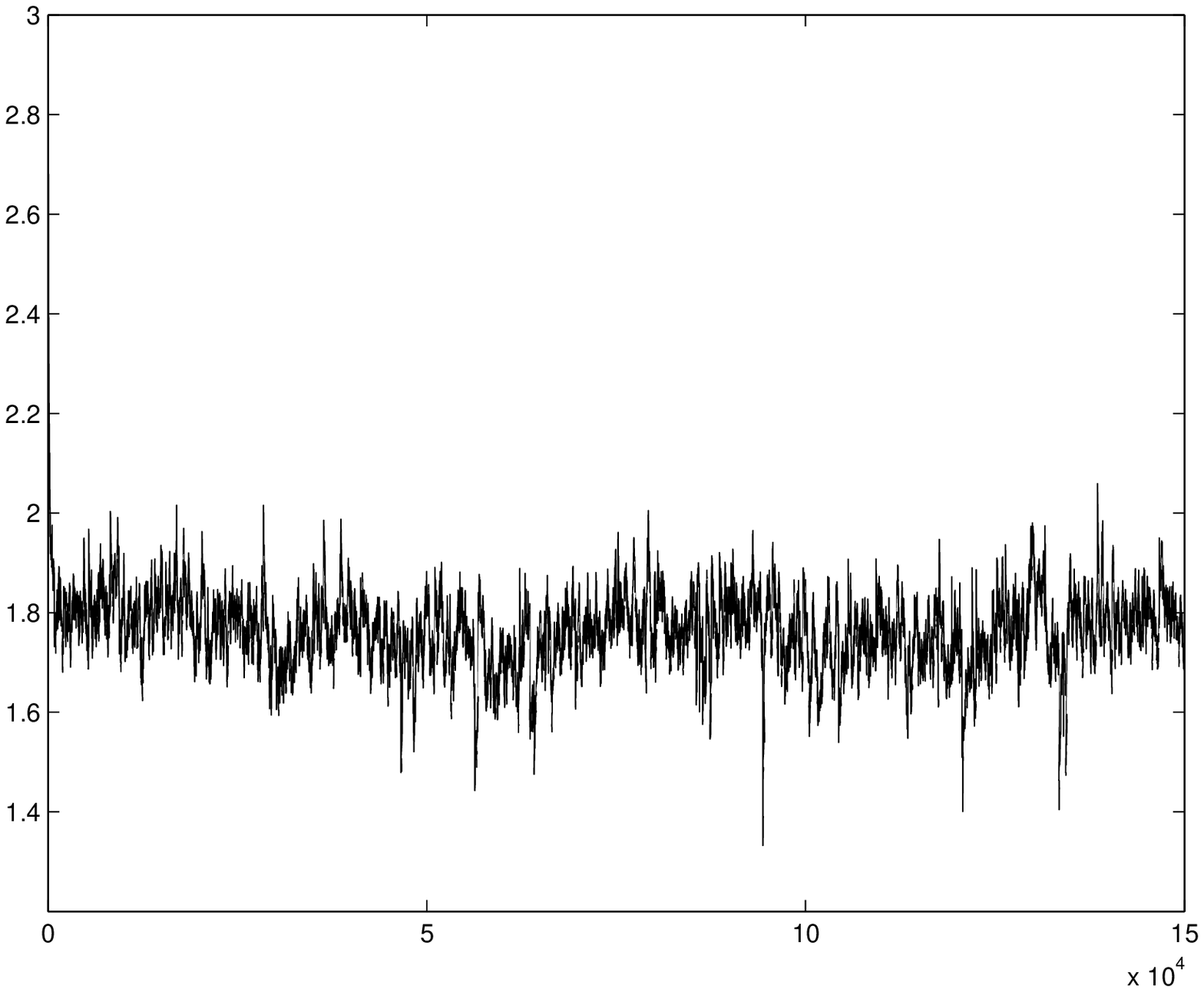}\\
\multicolumn{2}{c}{$\beta = 1.81$ - $\mathrm{PSRF} = 0.98$} \\
\includegraphics[width=8cm, height=3.4cm]{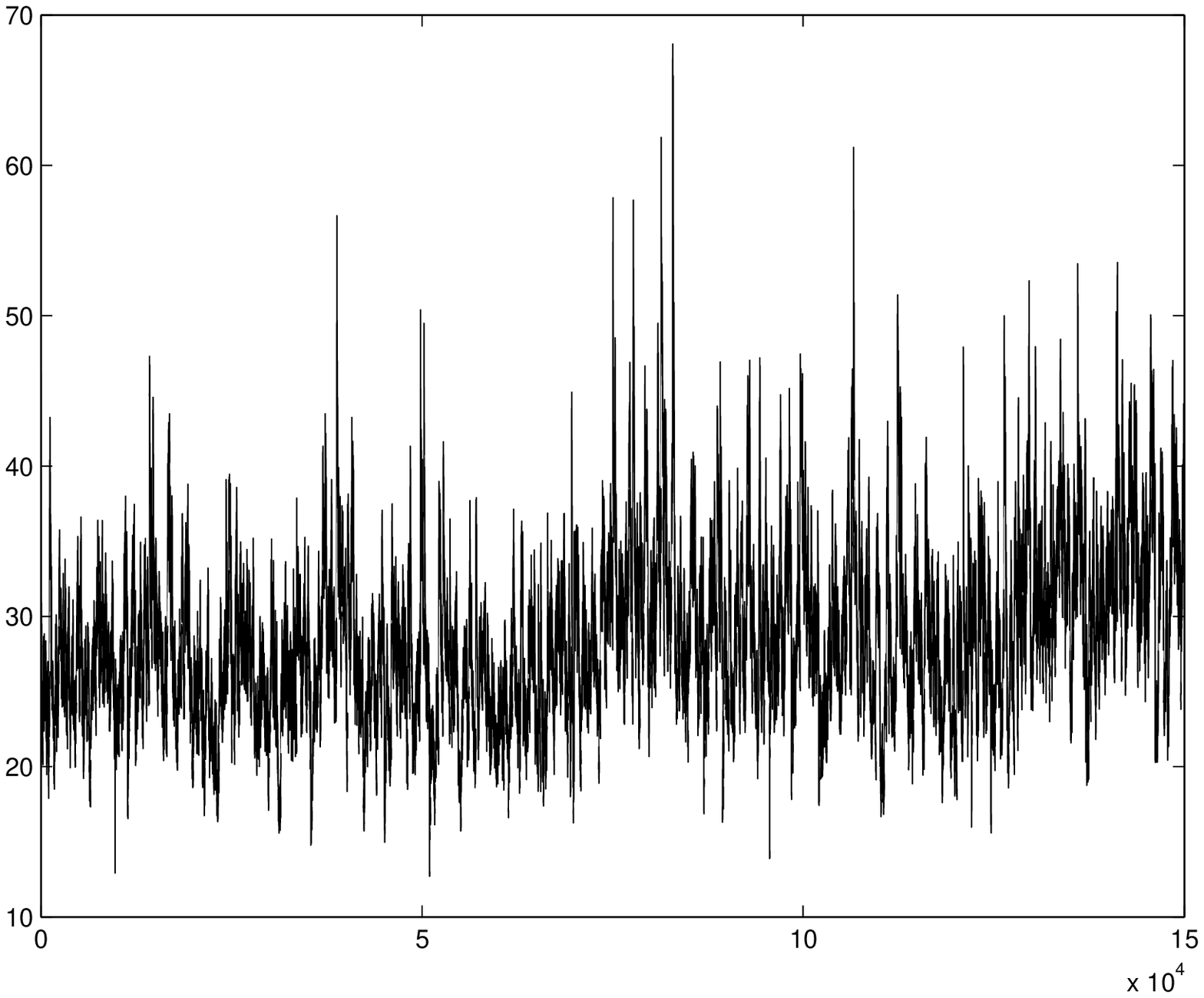}&
\includegraphics[width=8cm, height=3.4cm]{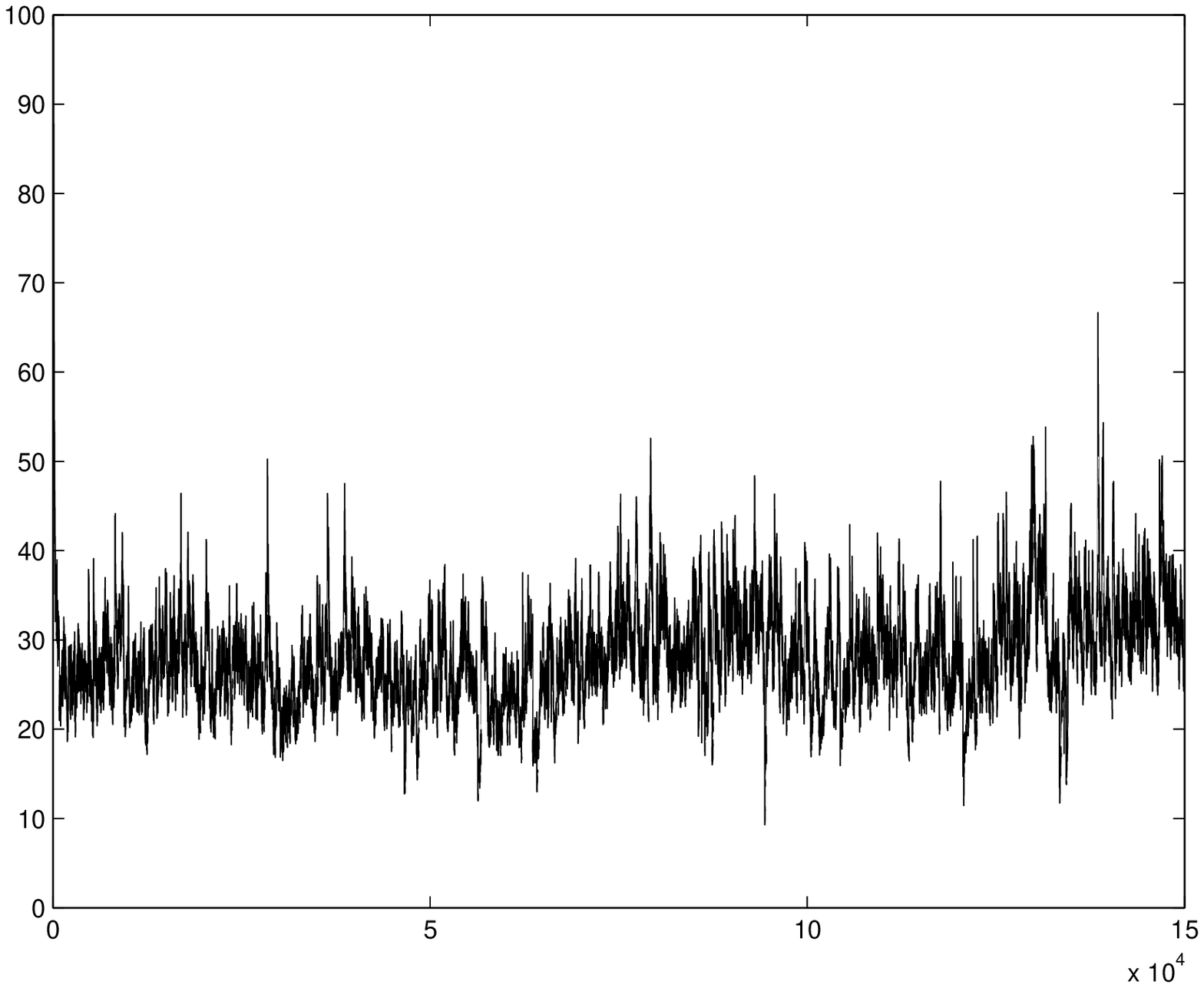} \\
\multicolumn{2}{c}{$\gamma = 31.5$ - $\mathrm{PSRF} = 1.03$} \\
\end{tabular}
\caption{Ground truth values and sample path for the hyper-parameters $\beta$ and $\gamma$
related to $v_{2,2}$ in $\mathcal{B}_2$.} \label{fig:cv2}
\end{figure}

To be able to automatically stop the simulated chain and ensure that the
last simulated samples are appropriately distributed according to the posterior
distribution of interest, a convergence monitoring technique based
on the potential scale reduction factor (PSRF) has been used by
simulating several chains in parallel (see \cite{gelman_92} for more
details). Using the union of two orthonormal bases as a frame representation, Figs.~\ref{fig:cv1} and \ref{fig:cv2}  show examples of convergence profiles corresponding to the hyper-parameters $\beta$ and $\gamma$ when two chains are sampled in parallel using Sampler 2.

Based on these values of the PSRF,
the algorithm was stopped after about $150,000$ iterations (burn-in period of $100,000$ iterations), which corresponds to about $4$ hours of computational time using Matlab~7.7
on an Intel Core 4 ($3$ GHz) architecture. 
When comparing the two proposed samplers in terms of convergence speed, it turns out from our simulations that Sampler 1 shows faster 
convergence than Sampler 2. Indeed, Sampler 1 needs about $110,000$ 
iterations to converge, which reduces the global computational time to 
about 3 hours.

\newpage

\begin{figure}[!ht]
\centering
\begin{tabular}{c c}
\centering
$\mathcal{B}_1$ & $\mathcal{B}_2$\\
\includegraphics[width=8cm, height=5cm]{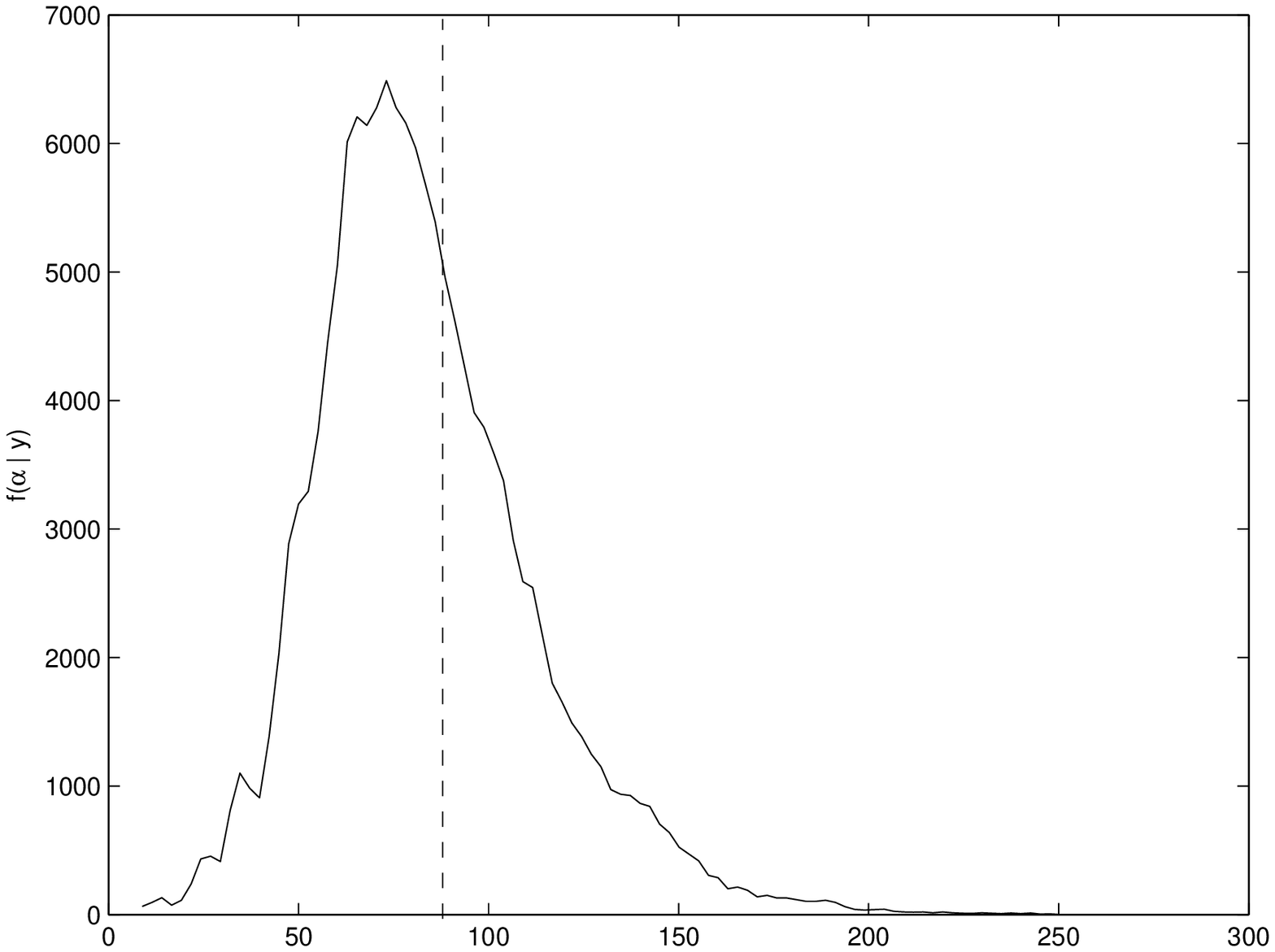} &
\includegraphics[width=8cm, height=5cm]{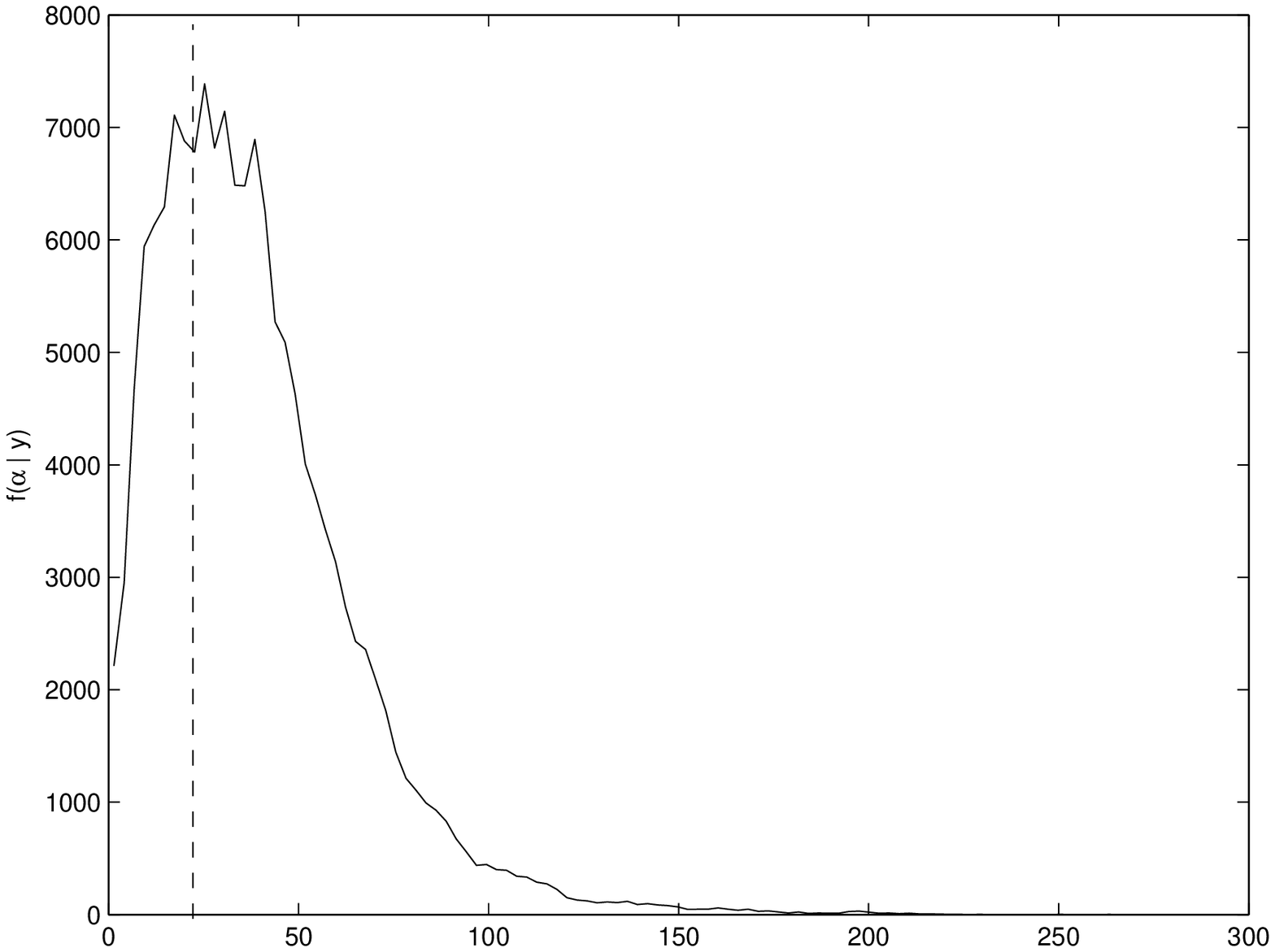}\\
$\gamma = 85.5$ & $\gamma = 24.07$ \\
\includegraphics[width=8cm, height=5cm]{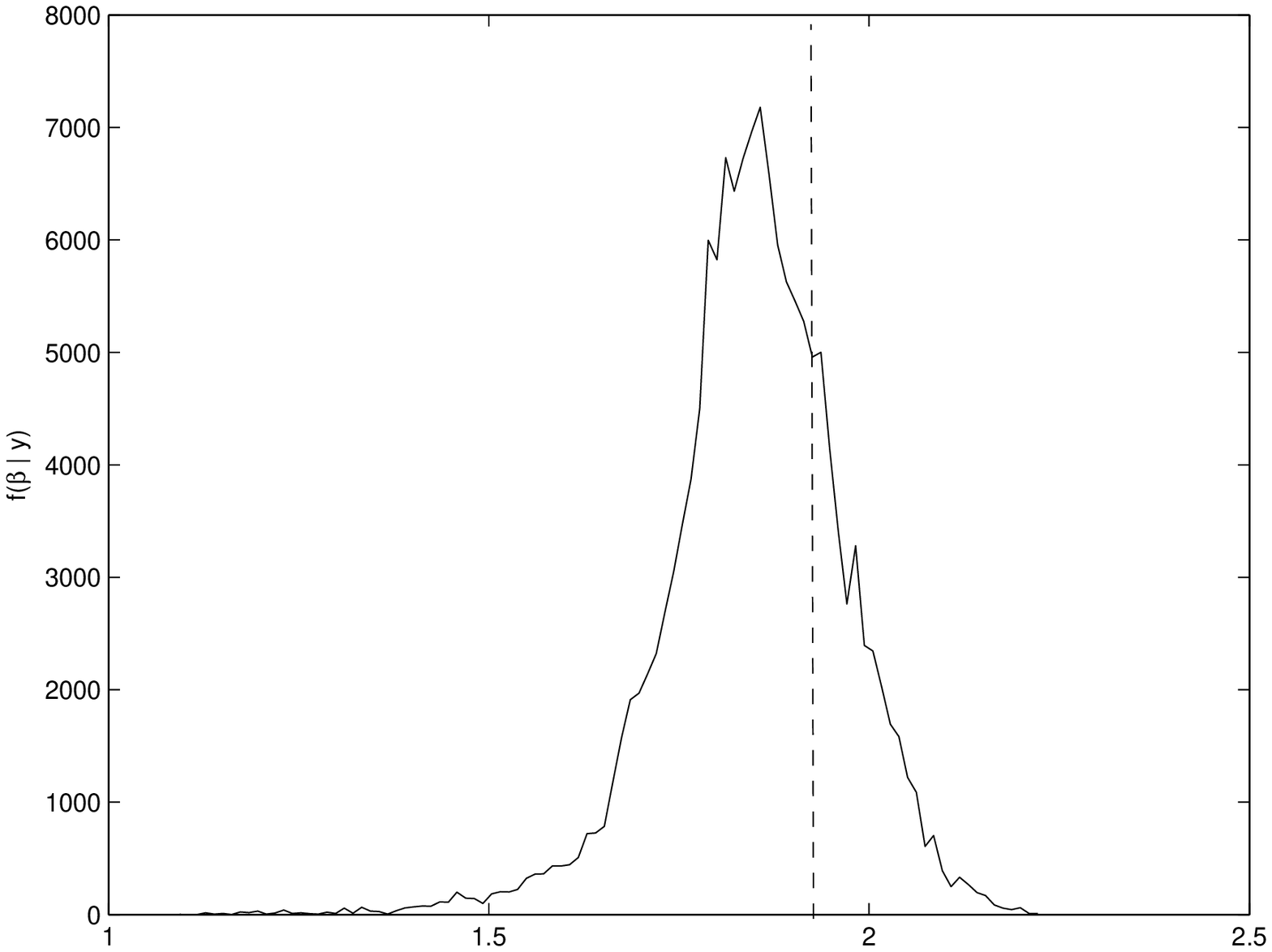}&
\includegraphics[width=8cm, height=5cm]{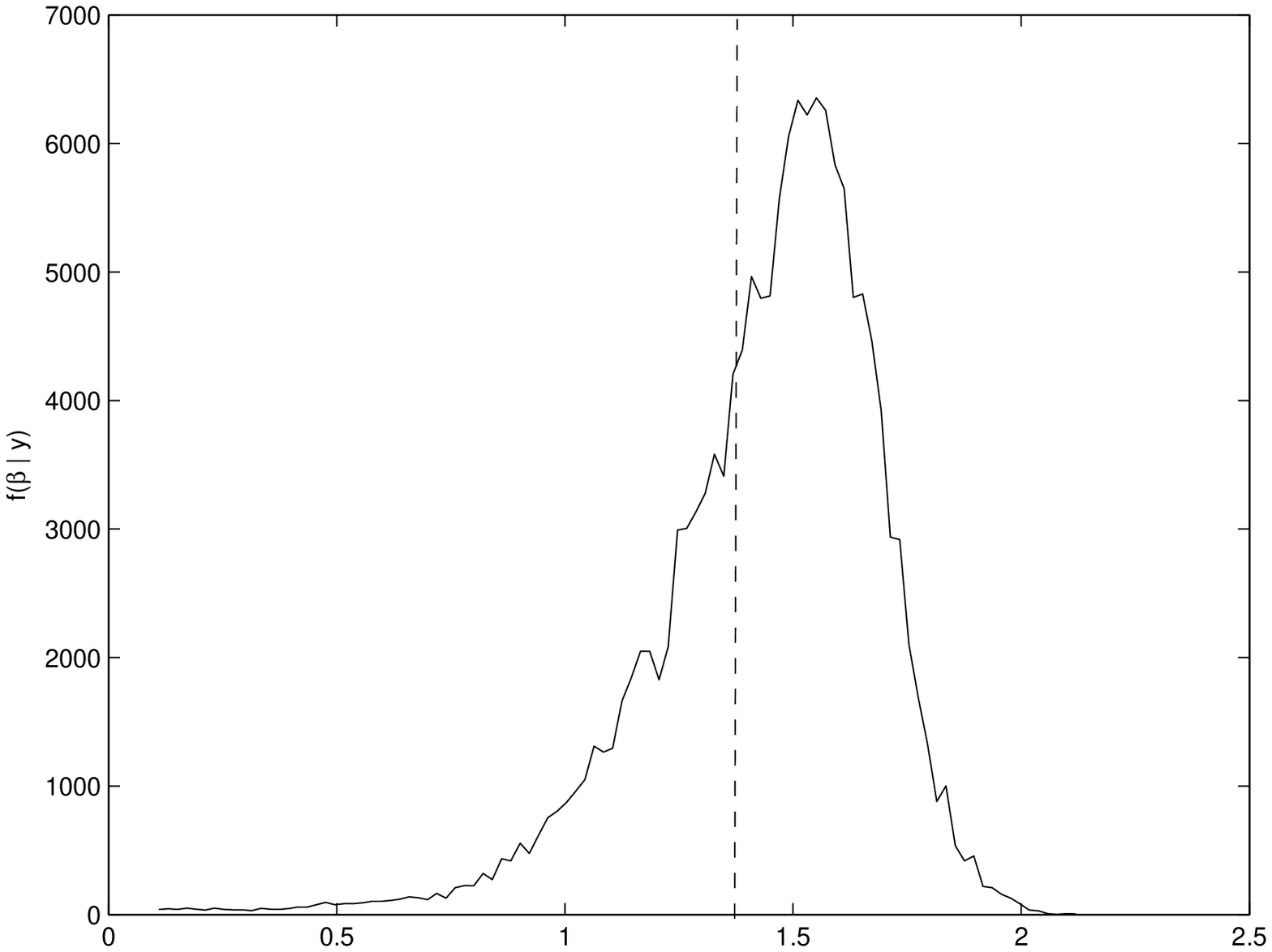} \\
$\beta = 1.87$ & $\beta = 1.35$ \\
\end{tabular}
\caption{Ground truth values (dashed line) and posterior distributions (solid line) of the sampled hyper-parameters
$\gamma$ and $\beta$,
for the subbands $h_{1,2}$ and $h_{2,2}$ in $\mathcal{B}_1$ and $\mathcal{B}_2$, repectively.} \label{fig:posteriors}
\end{figure}

The posterior distributions of the hyper-parameters $\beta$ and
$\gamma$ related to the subbands $h_{1,2}$ and $h_{2,2}$ in $\mathcal{B}_1$ and $\mathcal{B}_2$ introduced in Section~\ref{sec:Val_Ex1} are shown in Fig.~\ref{fig:posteriors}, as well as
the known original values. It is clear that the mode of the
posterior distributions is around
the ground truth value, which confirms the good estimation
performance of the proposed approach.

Note that when the resolution level increases, the number of
subbands also increases, which leads to a higher
number of hyper-parameters to be estimated and a potential increase of 
the required computational time to reach convergence.
For example, when using
the union of two orthonormal wavelet bases with two resolution
levels, the number of hyper-parameters to estimate is $28$. 
\newpage
\subsection{Application to image denoising}
\subsubsection{Example 1}
\begin{figure}[!ht]
\centering
\begin{tabular}{c c}
\centering
(a) &  (b)\\
\includegraphics[width=7cm, height=7cm]{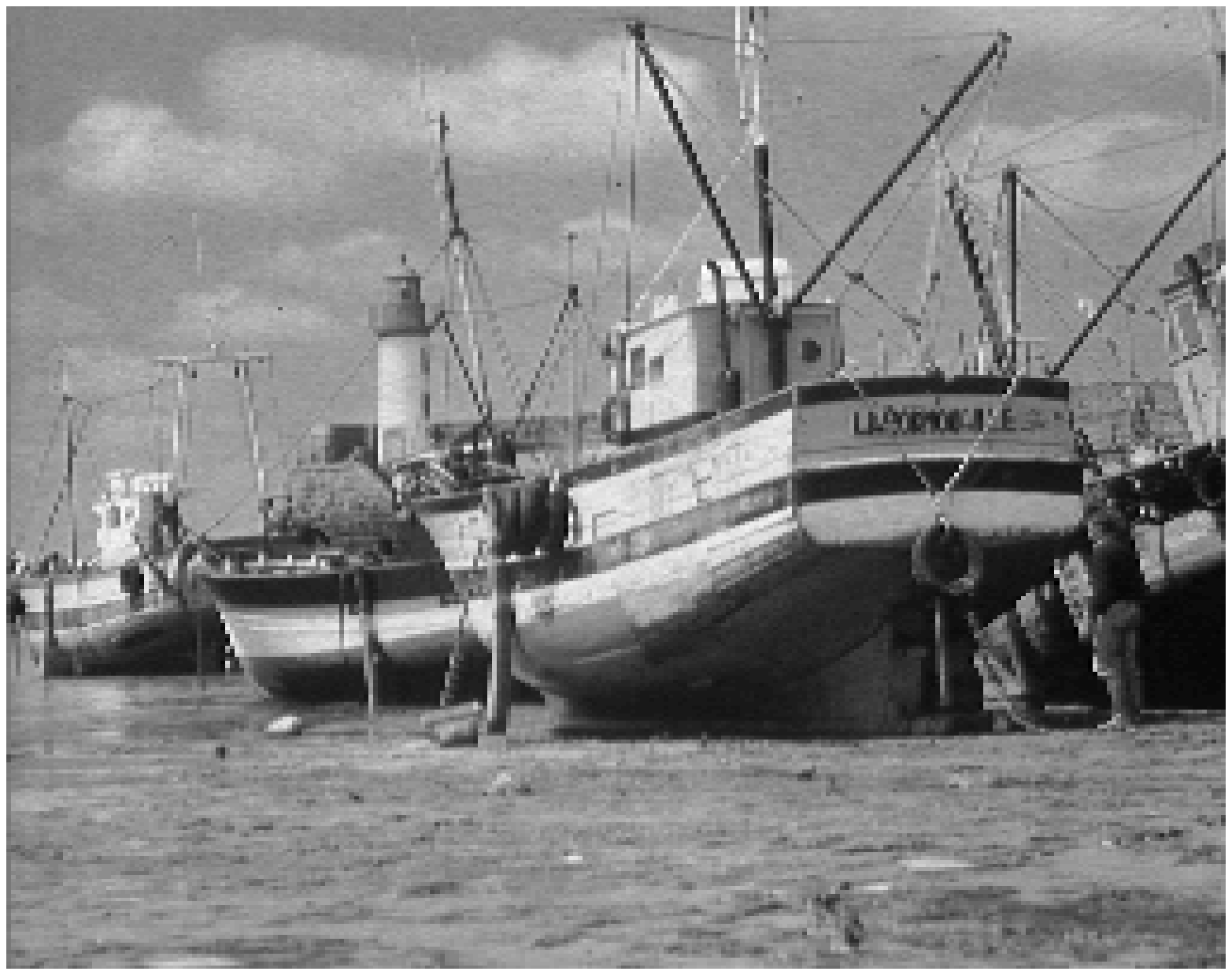} &
\includegraphics[width=7cm, height=7cm]{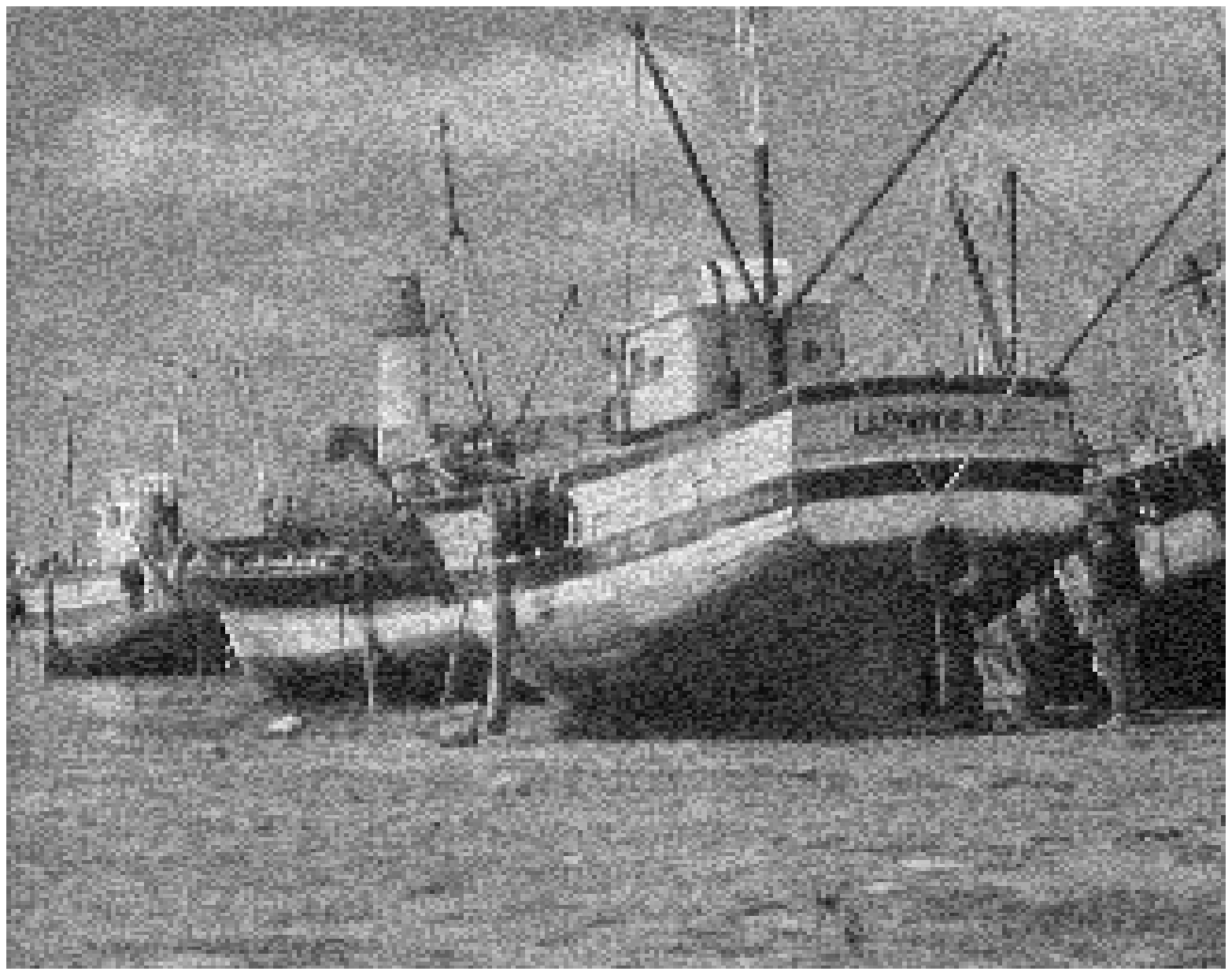}\\
(c) &(d) \\
\includegraphics[width=7cm, height=7cm]{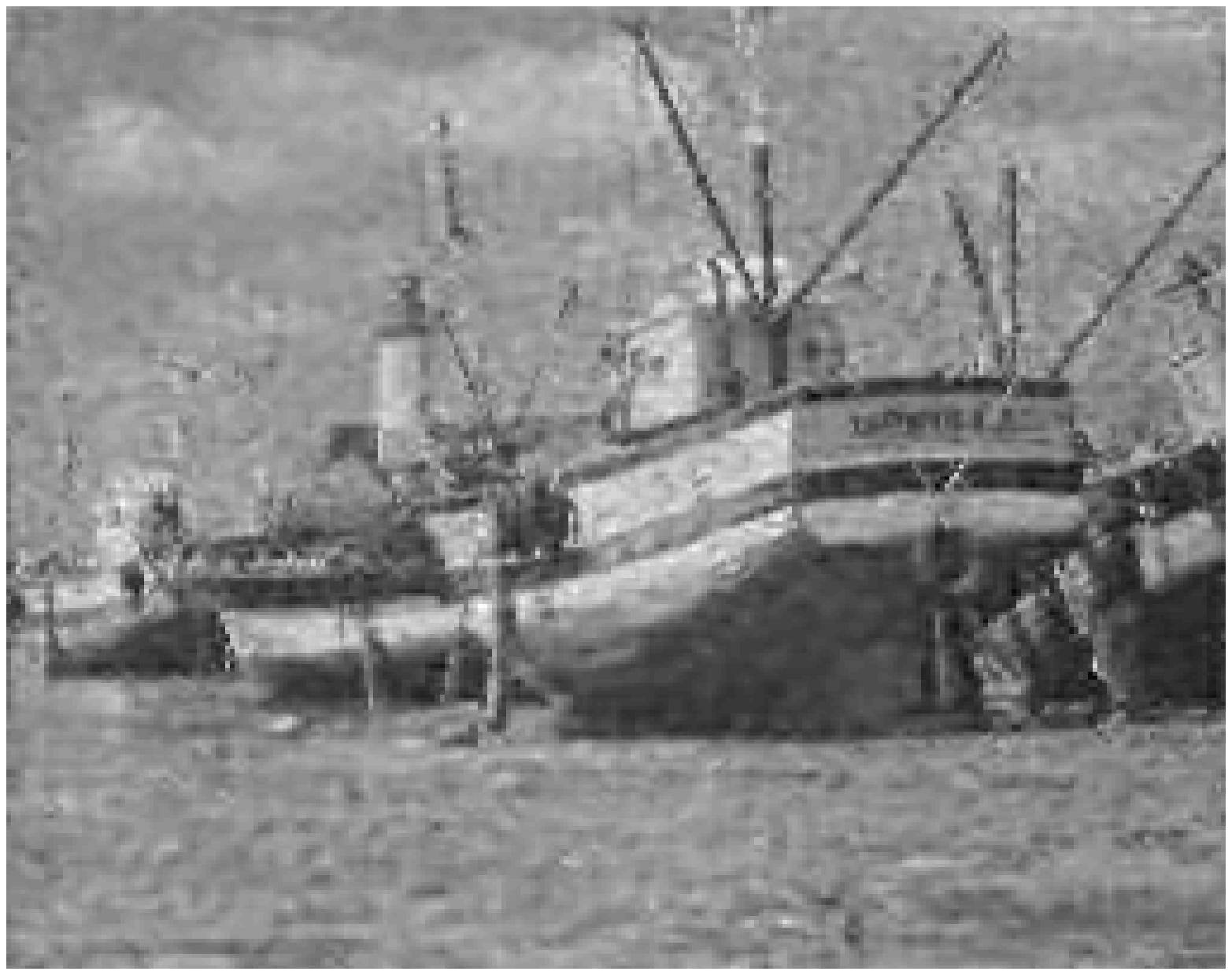}&
\includegraphics[width=7cm, height=7cm]{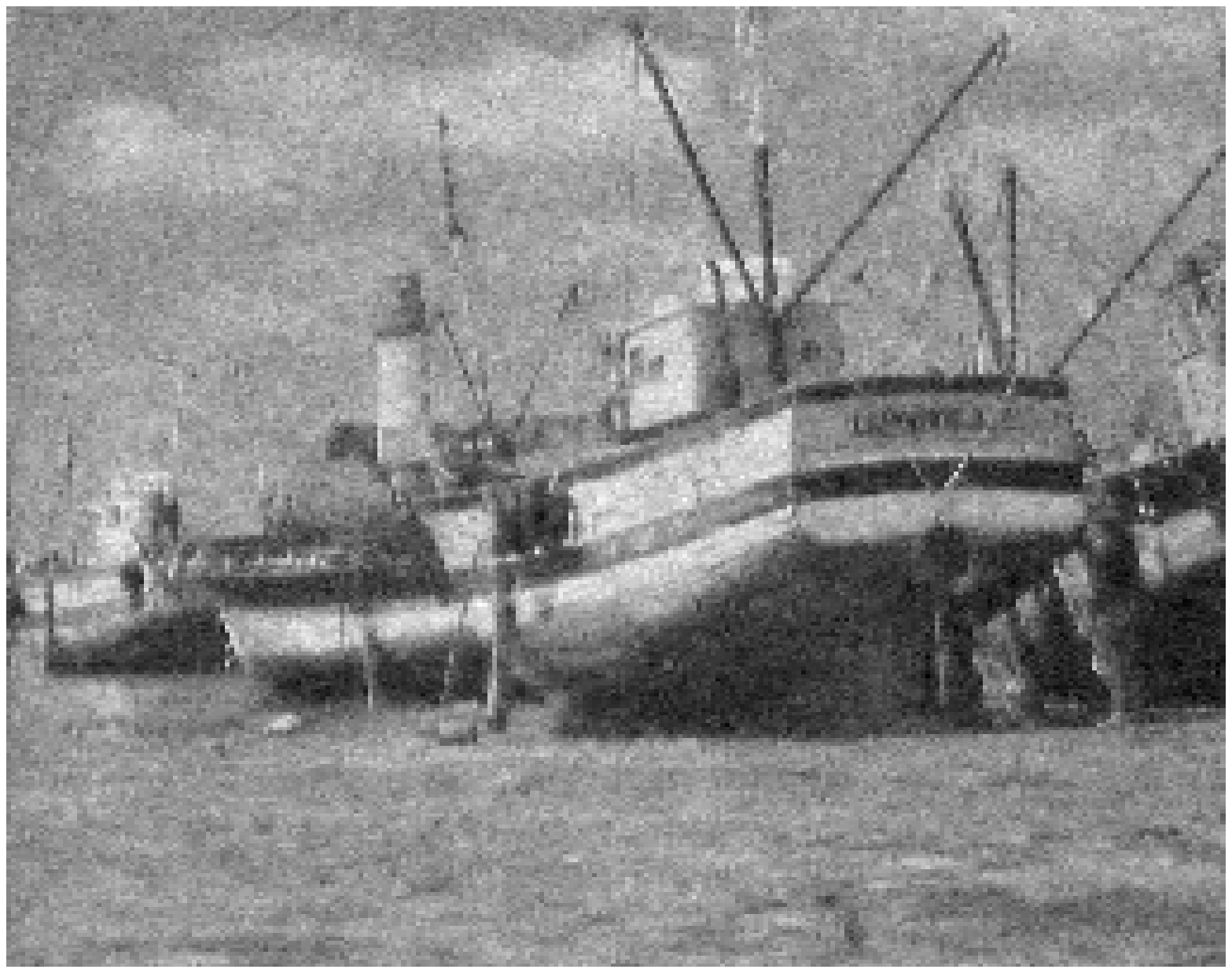} 
\end{tabular}
\caption{Original \emph{Boat} image (a), noisy image (b), denoised images using a variational approach (c) and the proposed MMSE estimator (d).} 
\label{fig:experiment1_boat}
\end{figure}

\hfill \\
\label{sec:exp1}
In this experiment, we are interested in recovering an image (the \textit{Boat} image of size
$256\times 256$) from its noisy observation affected by a noise
$\vect{n}$ uniformly distributed over the ball
$[-\delta,\delta]^{256 \times 256}$ with $\delta=30$.
We recall that the observation model for this image denoising
problem is given by \eqref{eq:delta}.
The noisy image in Fig.~\ref{fig:experiment1_boat} (b) was simulated using the available reference image $\vect{y}_{\rm ref}$ in Fig.~\ref{fig:experiment1_boat} (a) and the noise properties described above.\\
The union of two $2$D separable wavelet bases $\mathcal{B}_1$ and
$\mathcal{B}_2$ using Daubechies and shifted Daubechies filters of
length $8$ and $4$ (as for validation experiments in Section
\ref{sec:validation}) was used as a tight frame representation.
Denoising was performed using the MMSE denoted as
$\hat{\vect{x}}$ computed from sampled wavelet frame coefficients.
The adjoint frame operator is then applied to recover the denoised
image from its denoised estimated wavelet frame coefficients
($\hat{\vect{y}} = F^*\hat{\vect{x}}$). The obtained
denoised image is depicted in Fig.~\ref{fig:experiment1_boat} (d). For
comparison purpose, the denoised image using a variational approach \cite{chaux_07,combettes_pesquet_09} based on a MAP criterion using the estimated values of the hyper-parameters with our approach is illustrated in Fig.~\ref{fig:experiment1_boat}~(c).
This comparison shows that, for denoising purposes, the proposed method gives better visual quality than the other reported methods.\\

\begin{figure}[!ht]
\centering
\begin{tabular}{c c}
\centering
(a) &  (b)\\
\includegraphics[width=7cm, height=7cm]{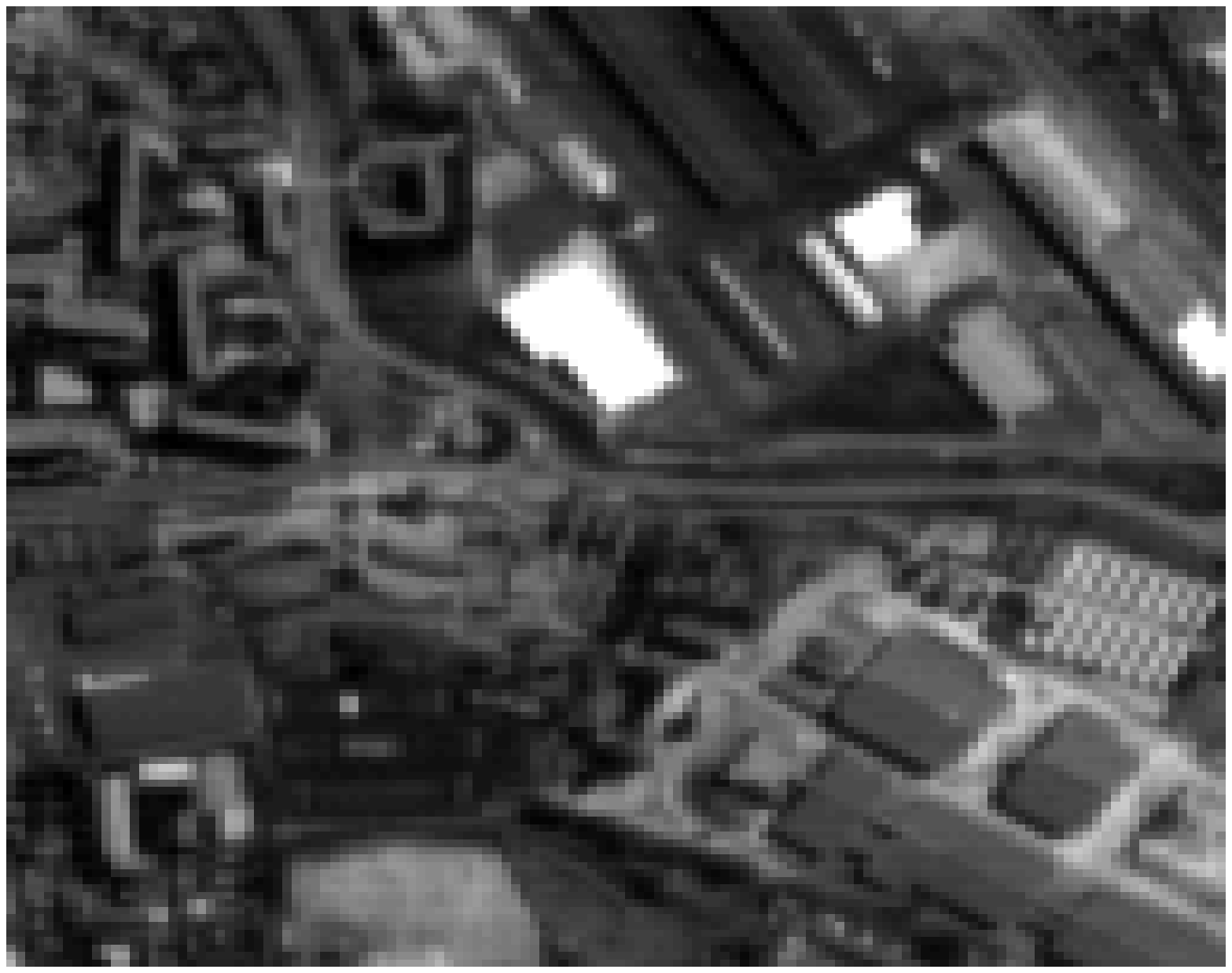} &
\includegraphics[width=7cm, height=7cm]{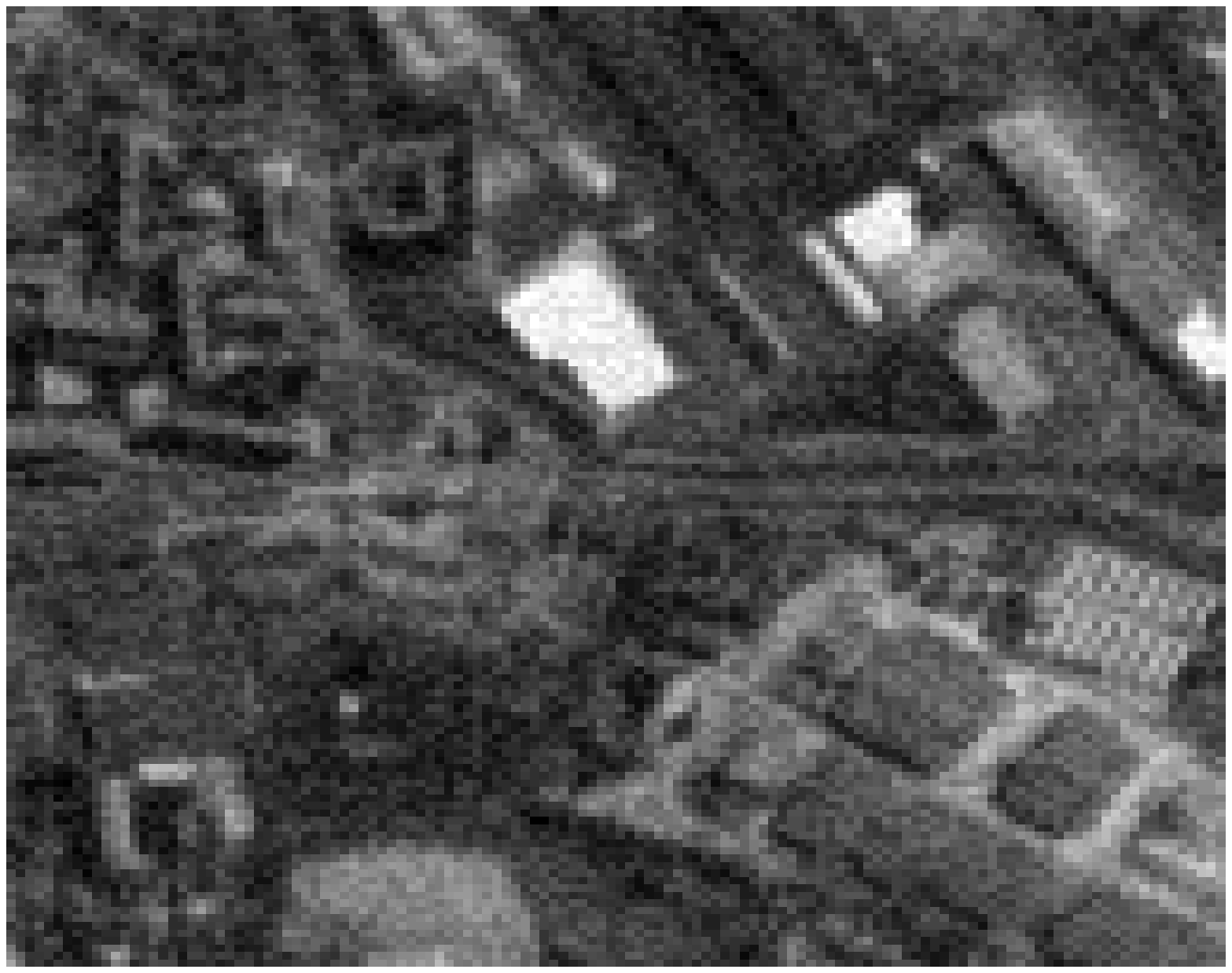}\\
(c) &(d) \\
\includegraphics[width=7cm, height=7cm]{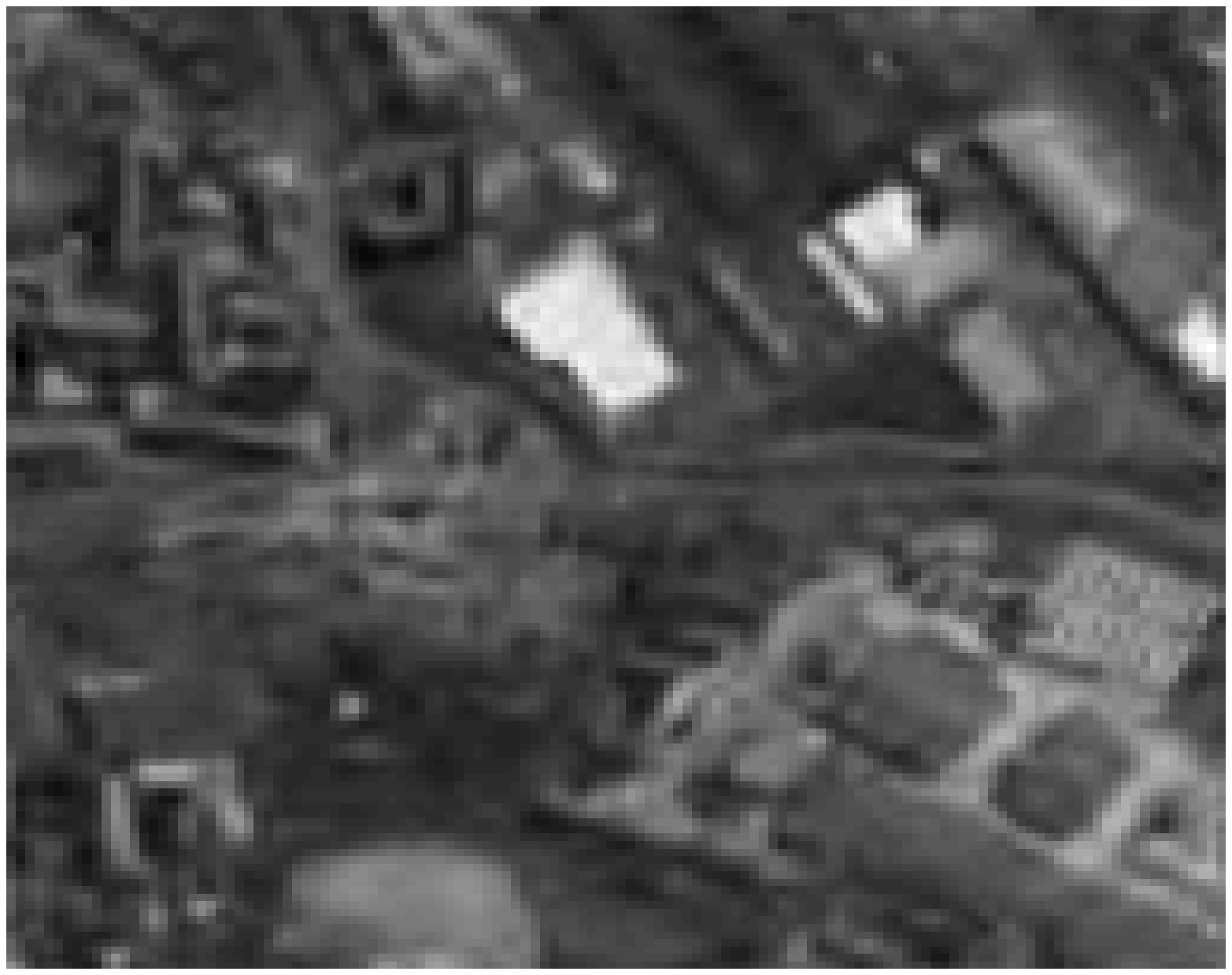}&
\includegraphics[width=7cm, height=7cm]{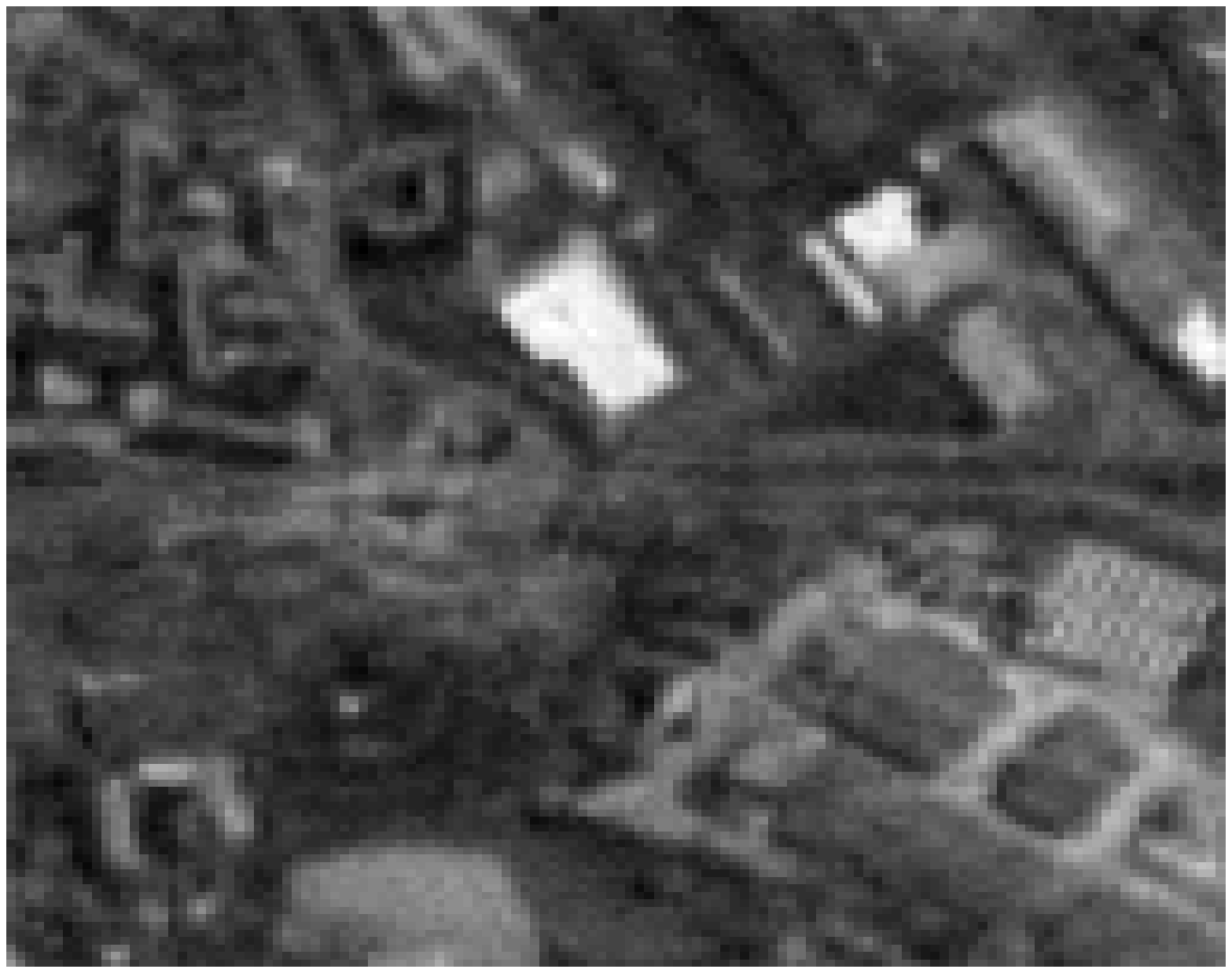} 
\end{tabular}
\caption{Original \emph{Marseille} image (a), noisy image (b), denoised images using a variational approach (c) and the proposed MMSE estimator (d).} 
\label{fig:experiment1_marseille}
\end{figure}

\begin{figure}[!ht]
\centering
\begin{tabular}{c c}
\centering
(a) &  (b)\\
\includegraphics[width=7cm, height=7cm]{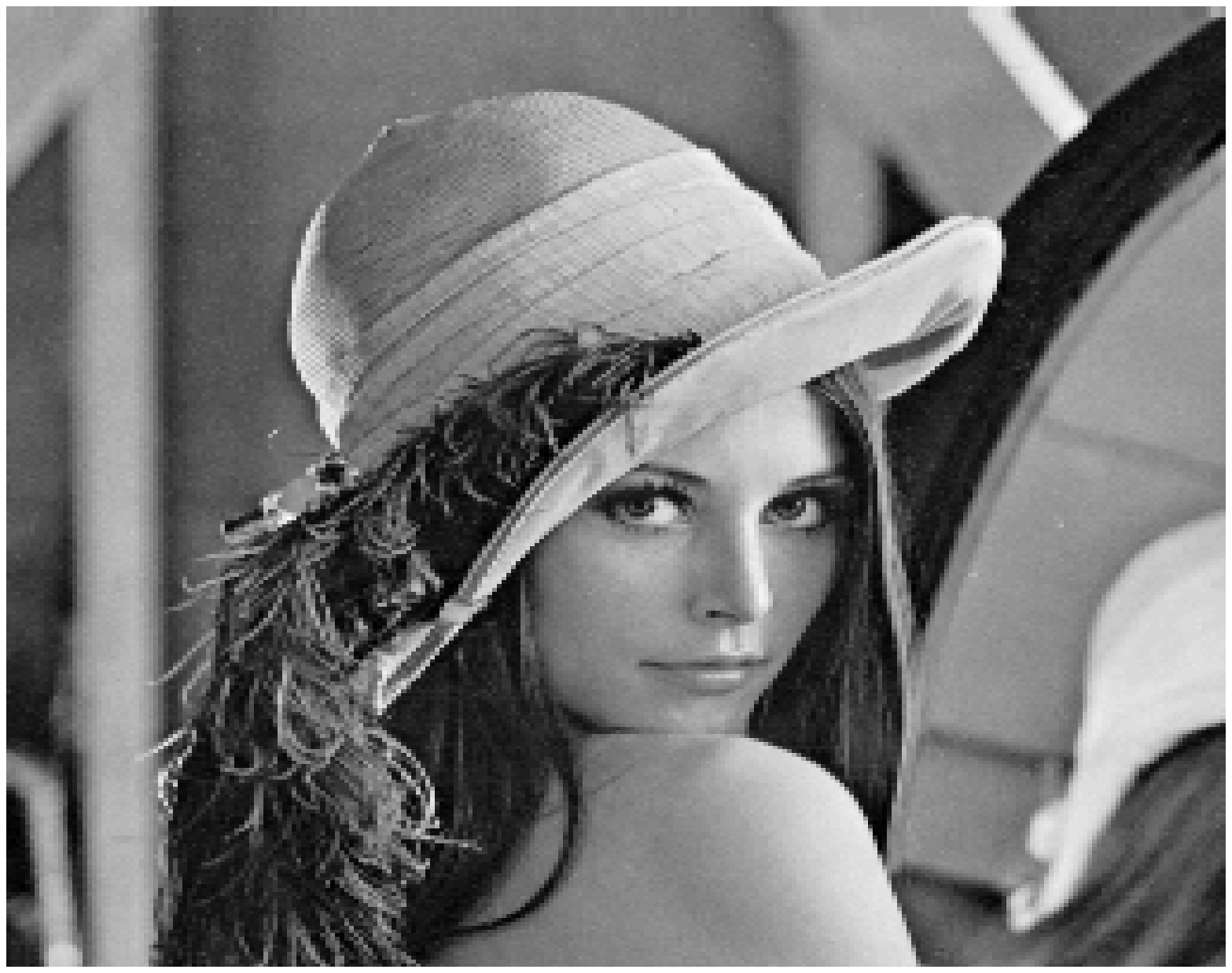} &
\includegraphics[width=7cm, height=7cm]{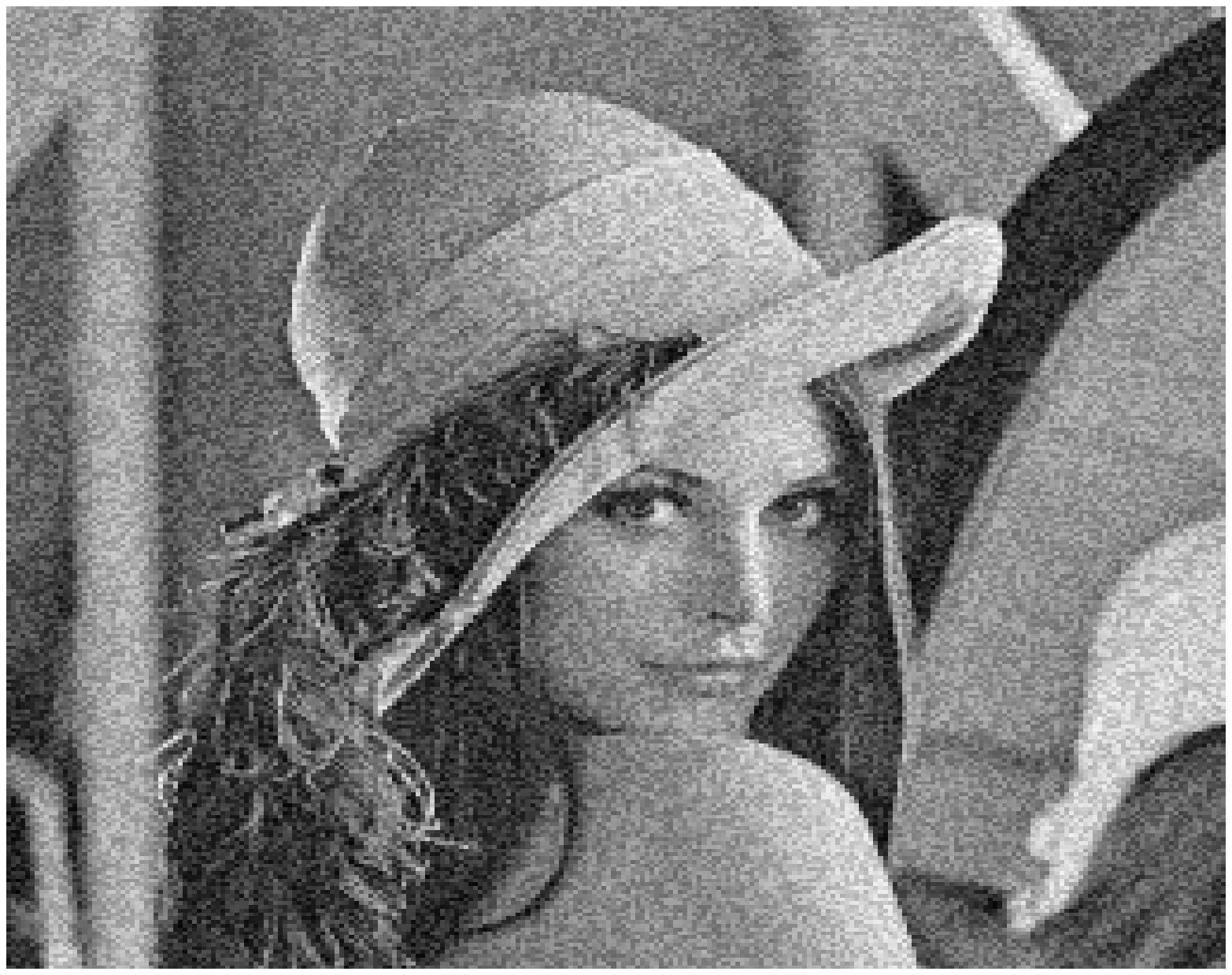}\\
(c) &(d) \\
\includegraphics[width=7cm, height=7cm]{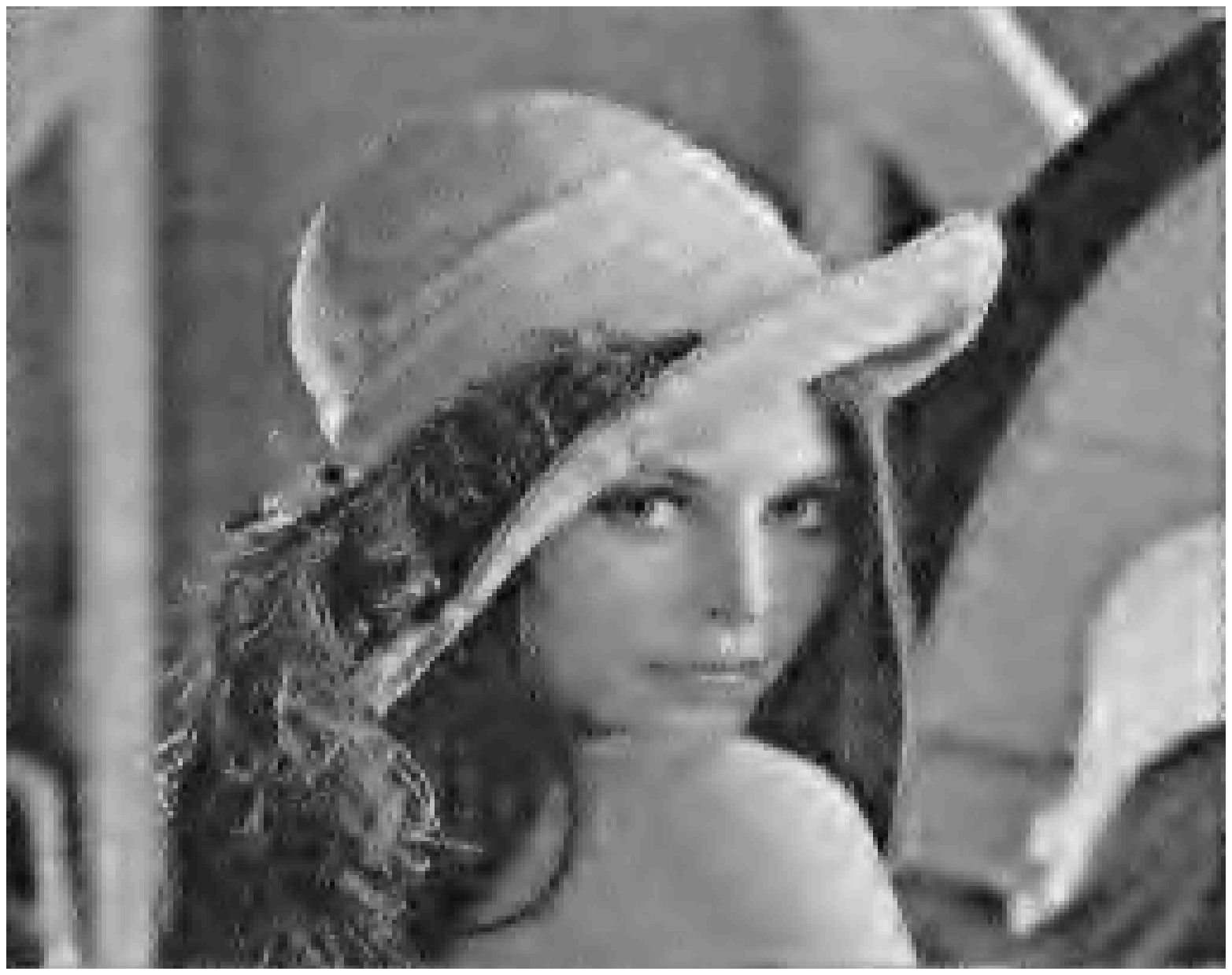}&
\includegraphics[width=7cm, height=7cm]{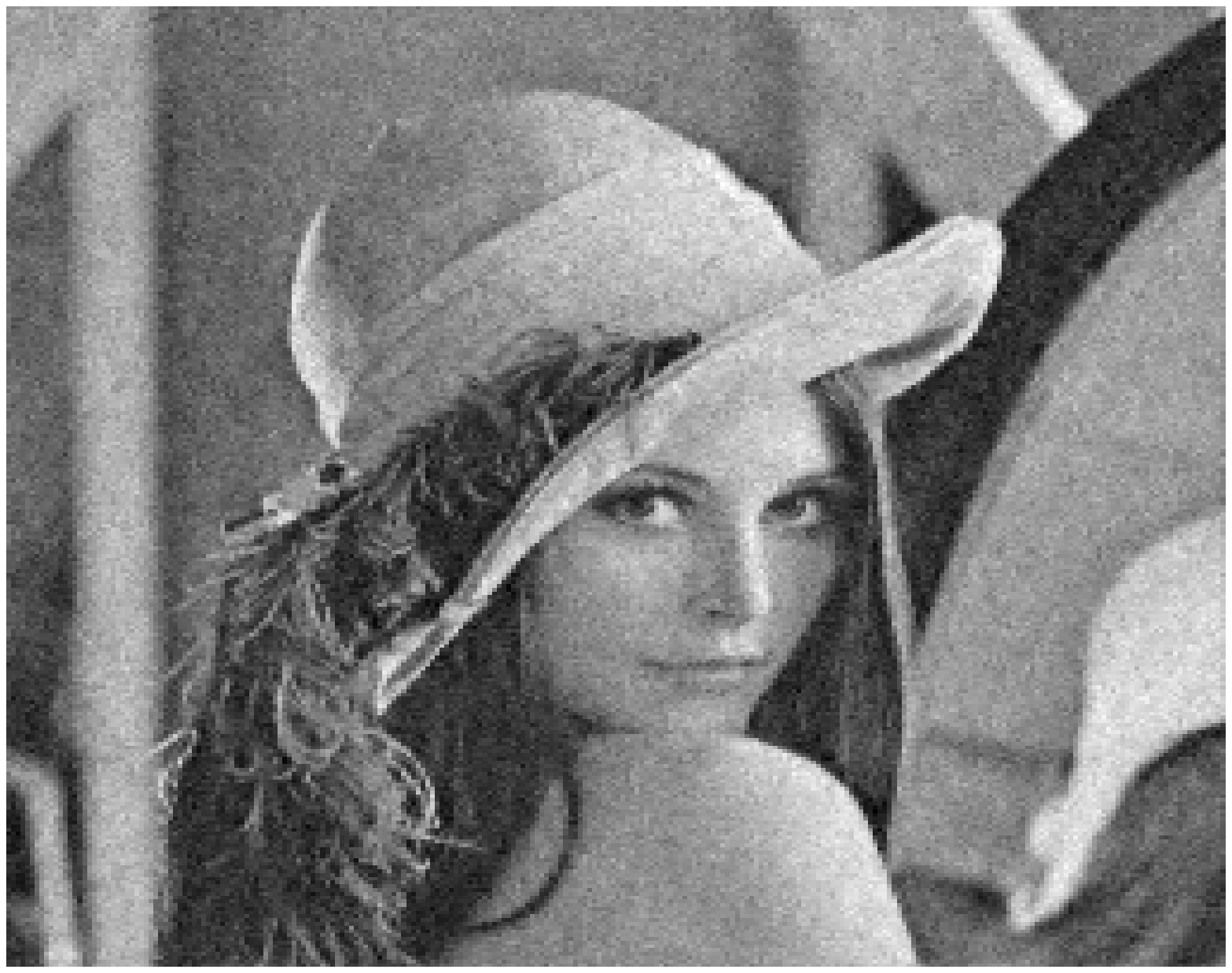} 
\end{tabular}
\caption{Original \emph{Lenna} image (a), noisy image (b), denoised images using a variational approach (c) and the proposed MMSE estimator (d).}
\label{fig:experiment1_lenna}
\end{figure}

\begin{figure}[!ht]
\centering
\begin{tabular}{c c}
\centering
(a) &  (b)\\
\includegraphics[width=7cm, height=7cm]{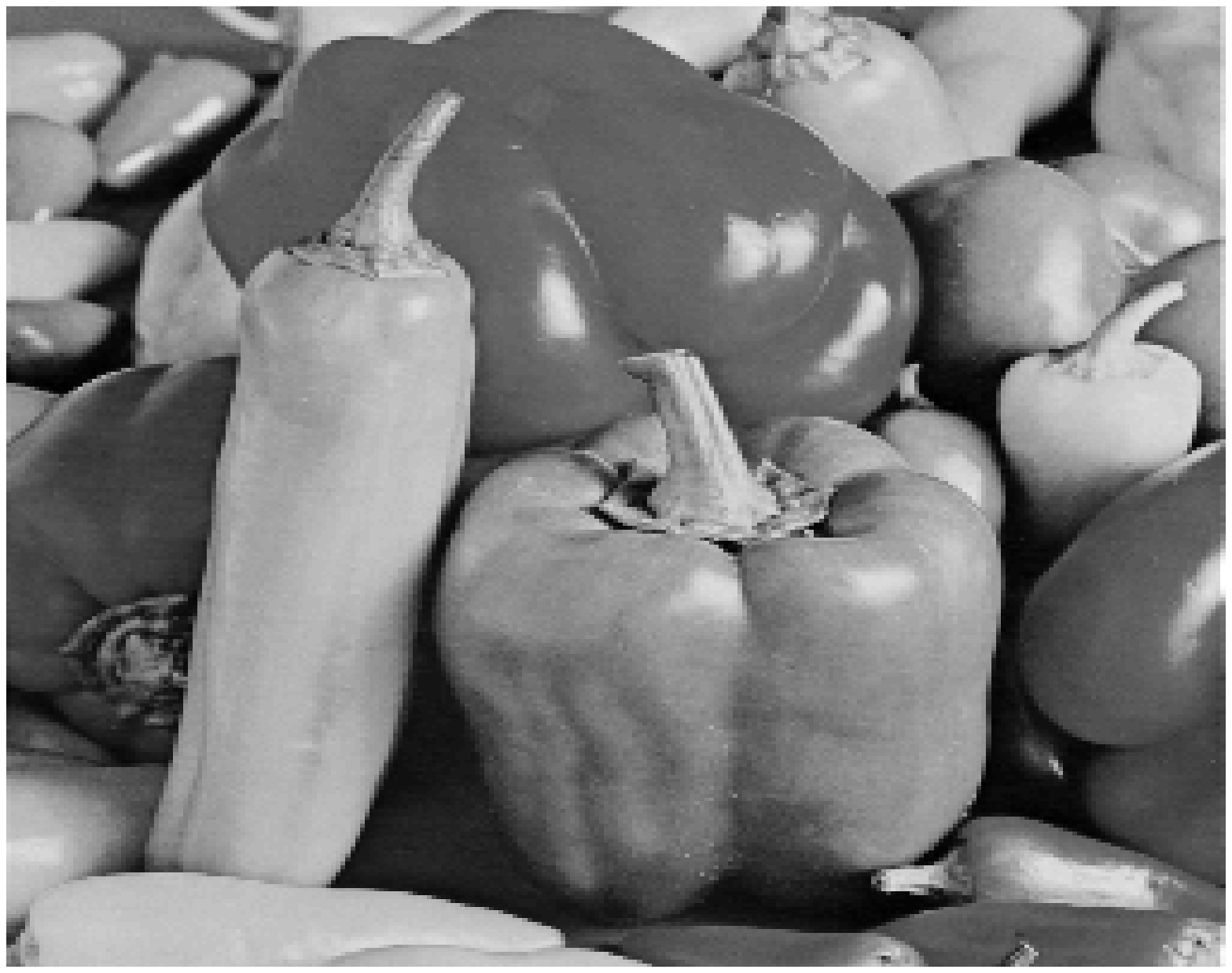} &
\includegraphics[width=7cm, height=7cm]{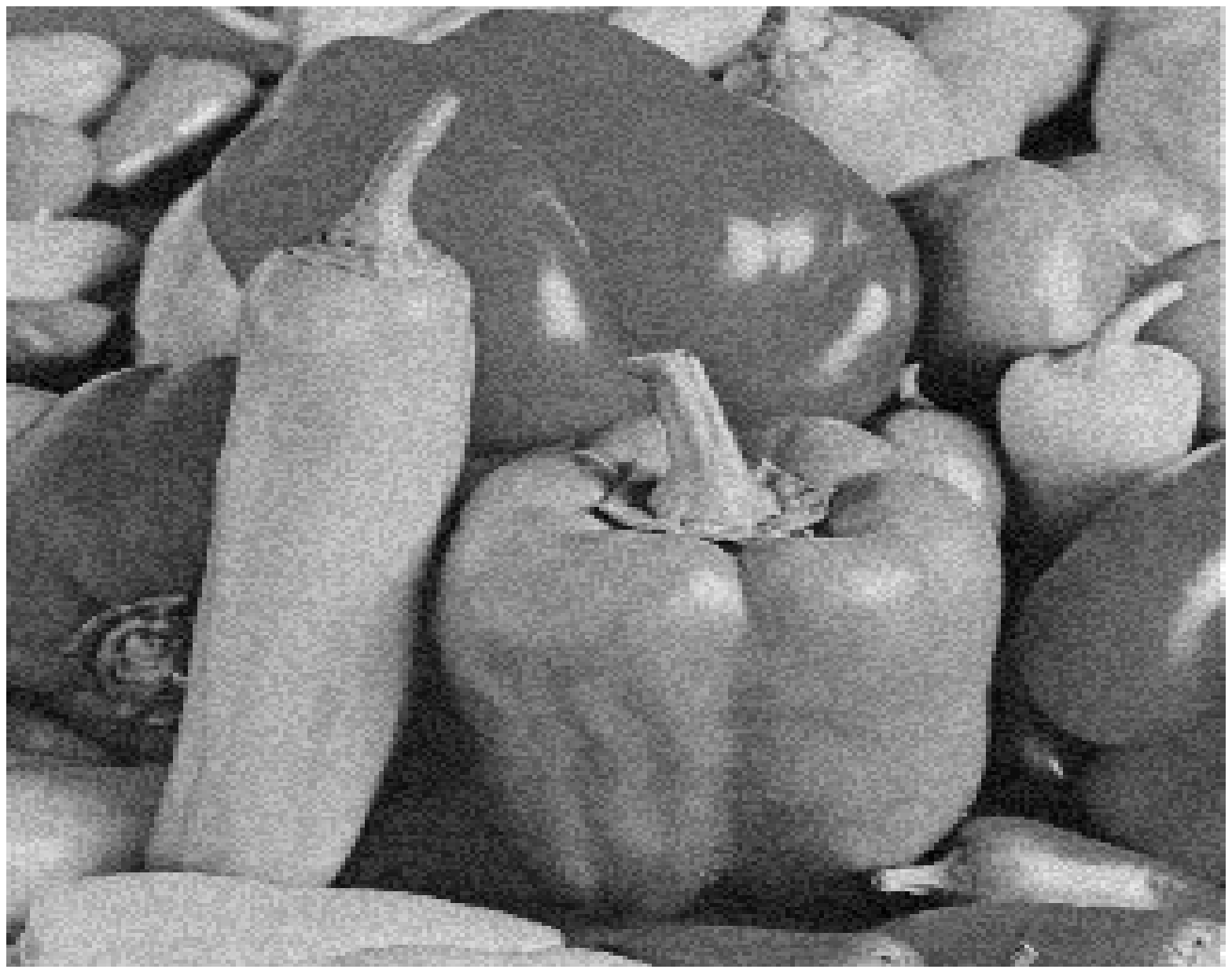}\\
(c) &(d) \\
\includegraphics[width=7cm, height=7cm]{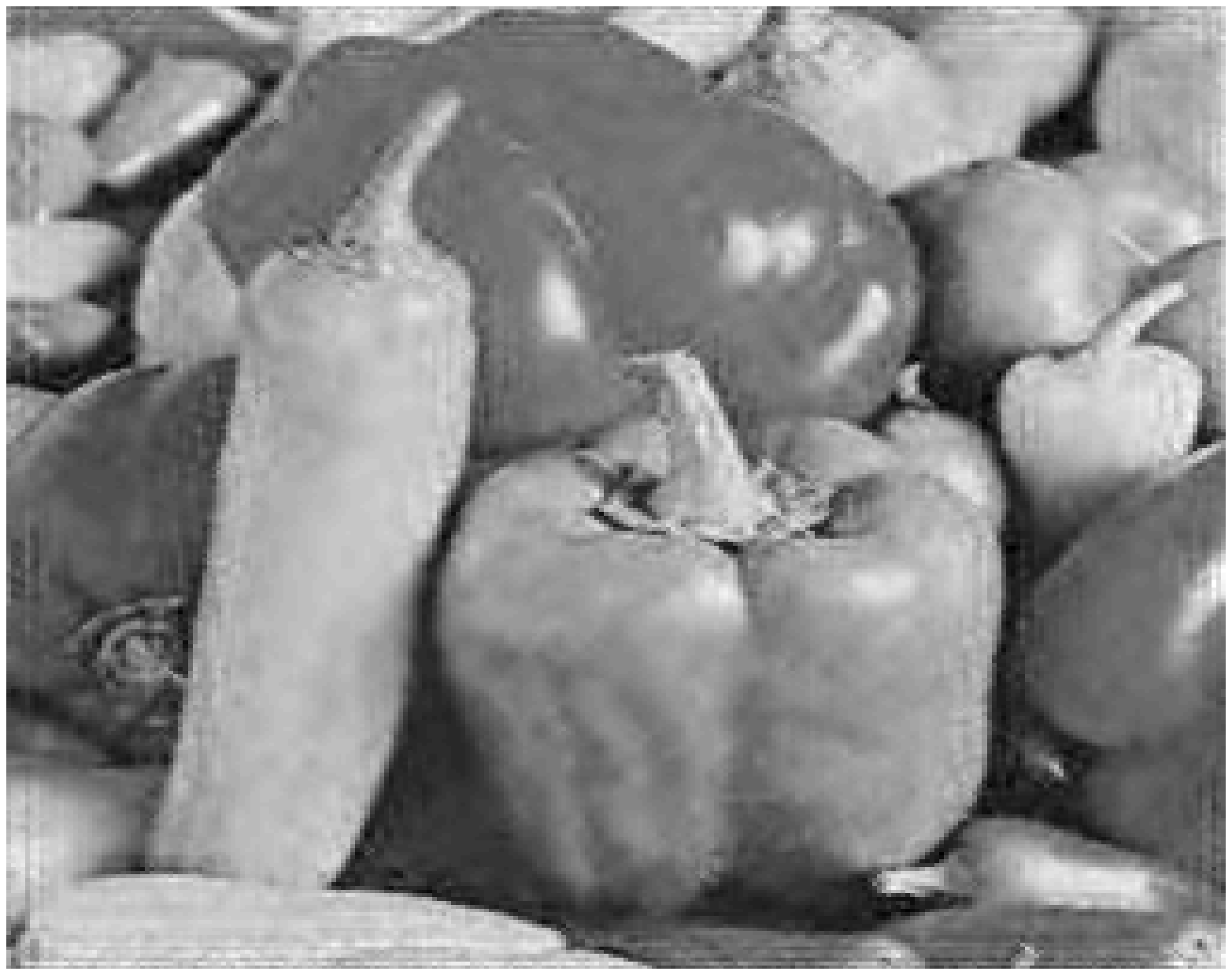}&
\includegraphics[width=7cm, height=7cm]{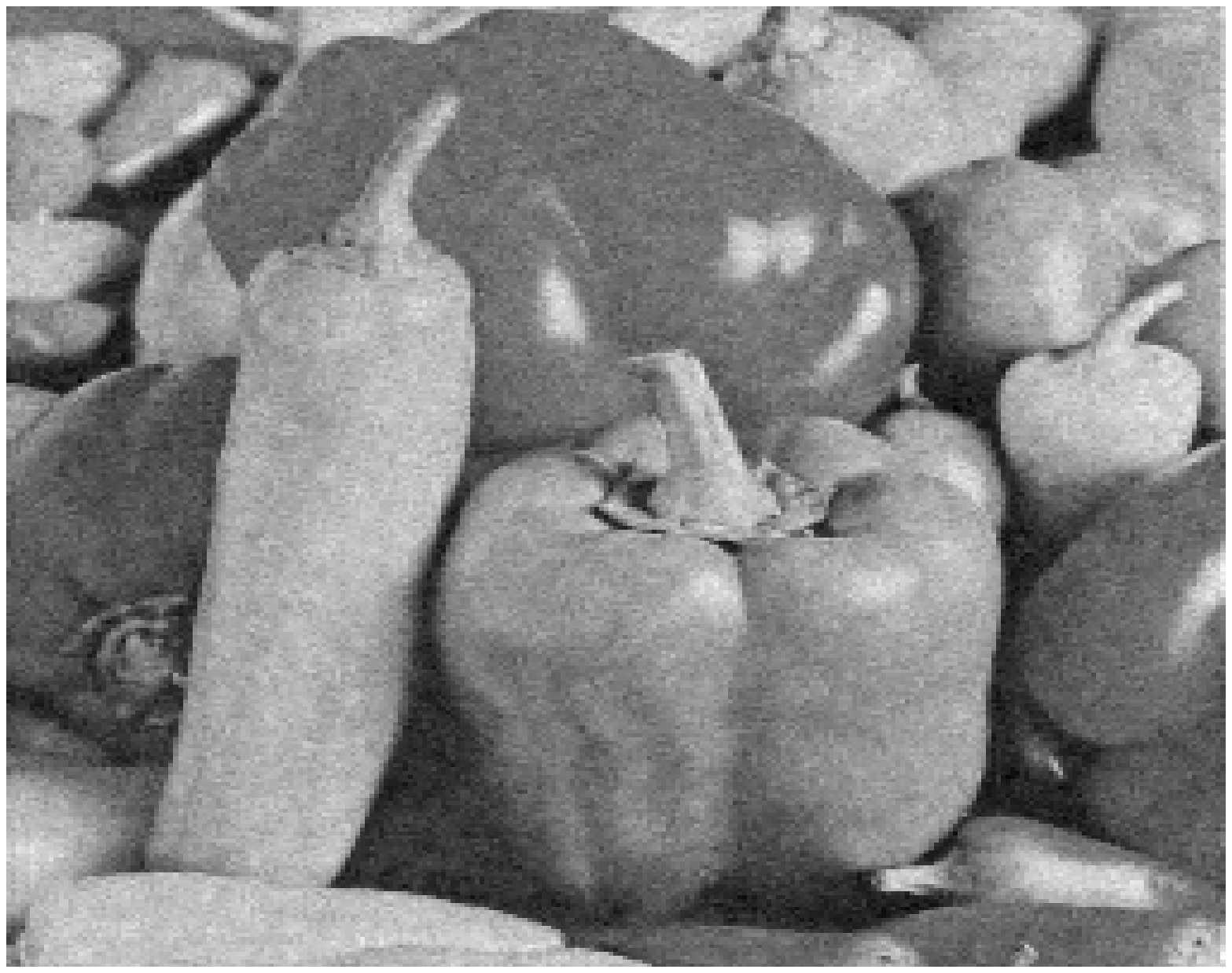} 
\end{tabular}
\caption{Original \emph{Peppers} image (a), noisy image (b), denoised images using a variational approach (c) and the proposed MMSE estimator (d).} 
\label{fig:experiment1_peppers}
\end{figure}

Signal to noise ratio ($\mathrm{SNR}=20\log_{10}\big(\| \vect{y}_{\rm ref}\|/\|
\vect{y}_{\rm ref}-\hat{\vect{y}}\|\big)$) and structural similarity (SSIM) \cite{Wang_04} values are also given in Table~\ref{tab:snr_ssim} to quantitatively evaluate denoising performance. Note here that SSIM values must lie in $[0,1]$, high values indicating good image quality.\\ 
An additional comparison with respect to Wiener filtering is given in this table. The SNR and SSIM values are given for three additional test images with different textures and contents to better illustrate the good performance of the proposed approach. The corresponding original, noisy and denoised images are displayed in Figs.~\ref{fig:experiment1_marseille}, \ref{fig:experiment1_lenna} and \ref{fig:experiment1_peppers}.

\begin{table}[!ht]
\centering \caption{SNR and SSIM values for the noisy and denoised images.}
\begin{tabular}{|l|c|c|c|c|c|c|}
\hline
\multicolumn{2}{|c|}{} &Noisy & Wiener & Variational & MCMC\\
\hline
&SNR (dB)& 16.67&18.02& 18.41&\textbf{19.20}\\
\cline{2-6}
\emph{Boat}&SSIM &0.521&0.553&0.570&\textbf{0.614}\\
\hline
&SNR (dB)& 18.53&19.27& 20.55&\textbf{20.77}\\
\cline{2-6}
\emph{Marseille}&SSIM &0.797&0.802&0.824&\textbf{0.866}\\
\hline
&SNR (dB)& 17.69&19.63&21.79&\textbf{22.13}\\
\cline{2-6}
\emph{Lenna}&SSIM &0.496&0.583&0.671&\textbf{0.695}\\
\hline
&SNR (dB)& 21.23&21.64& 22.40&\textbf{22.67}\\
\cline{2-6}
\emph{Peppers}&SSIM &0.754&0.781&0.807&\textbf{0.811}\\
\hline
\end{tabular}
\label{tab:snr_ssim}
\end{table}

It is worth noticing that the visual quality and quantitative results show that the denoised image based on the MMSE estimate of the wavelet frame coefficients is better than 
the one obtained with the Wiener filtering or the variational approach.
For the latter approach, it must be emphasized that the choice
of the hyper-parameters always constitutes a delicate problem, for which
our algorithm brings a numerical solution.
\newpage
\subsubsection{Example 2}
\begin{figure}[!ht]
\centering
\begin{tabular}{c c}
\centering
(a) &  (b)\\
\includegraphics[width=7cm, height=7cm]{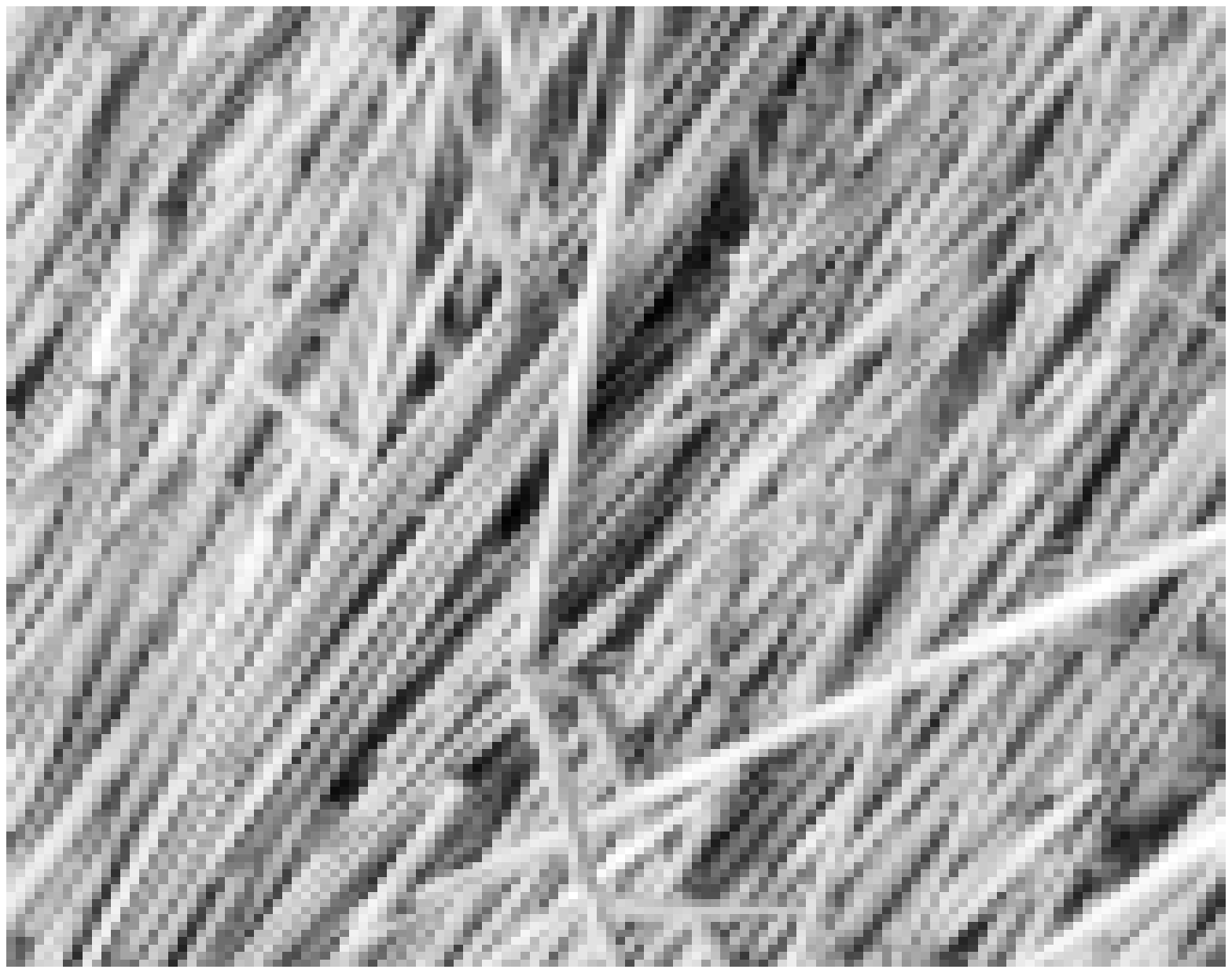} &
\includegraphics[width=7cm, height=7cm]{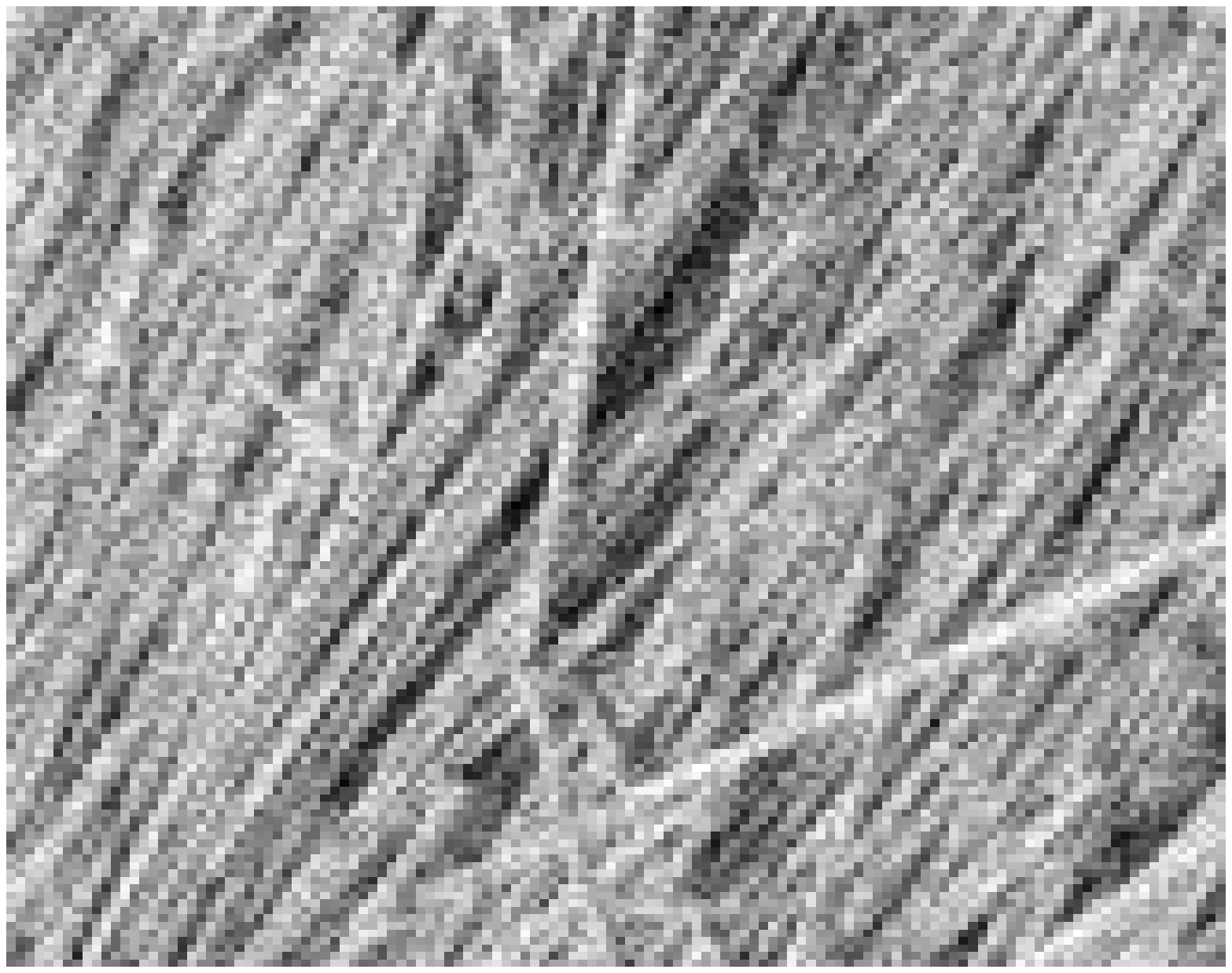}\\
(c)&(d)\\

\includegraphics[width=7cm, height=7cm]{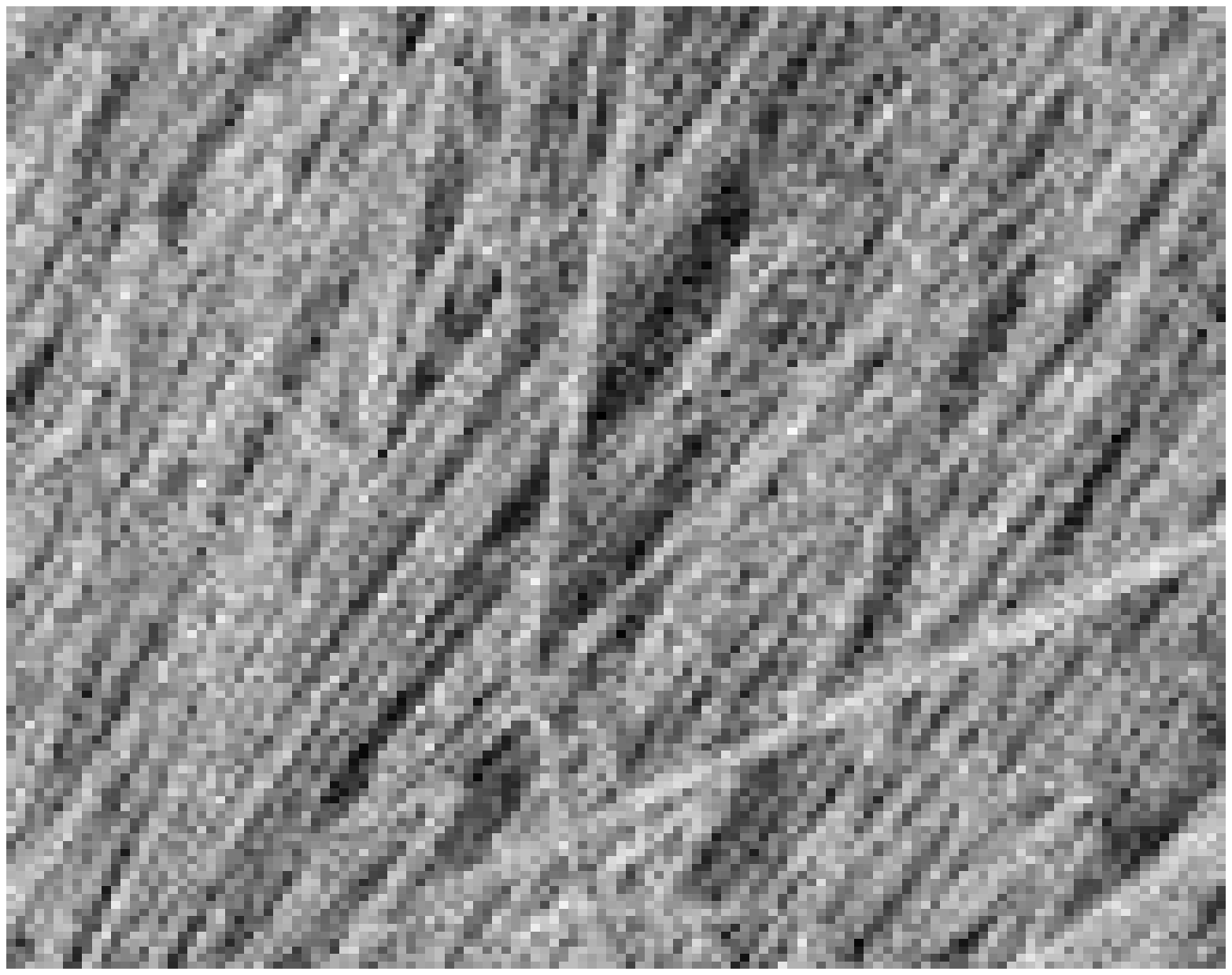}&
\includegraphics[width=7cm, height=7cm]{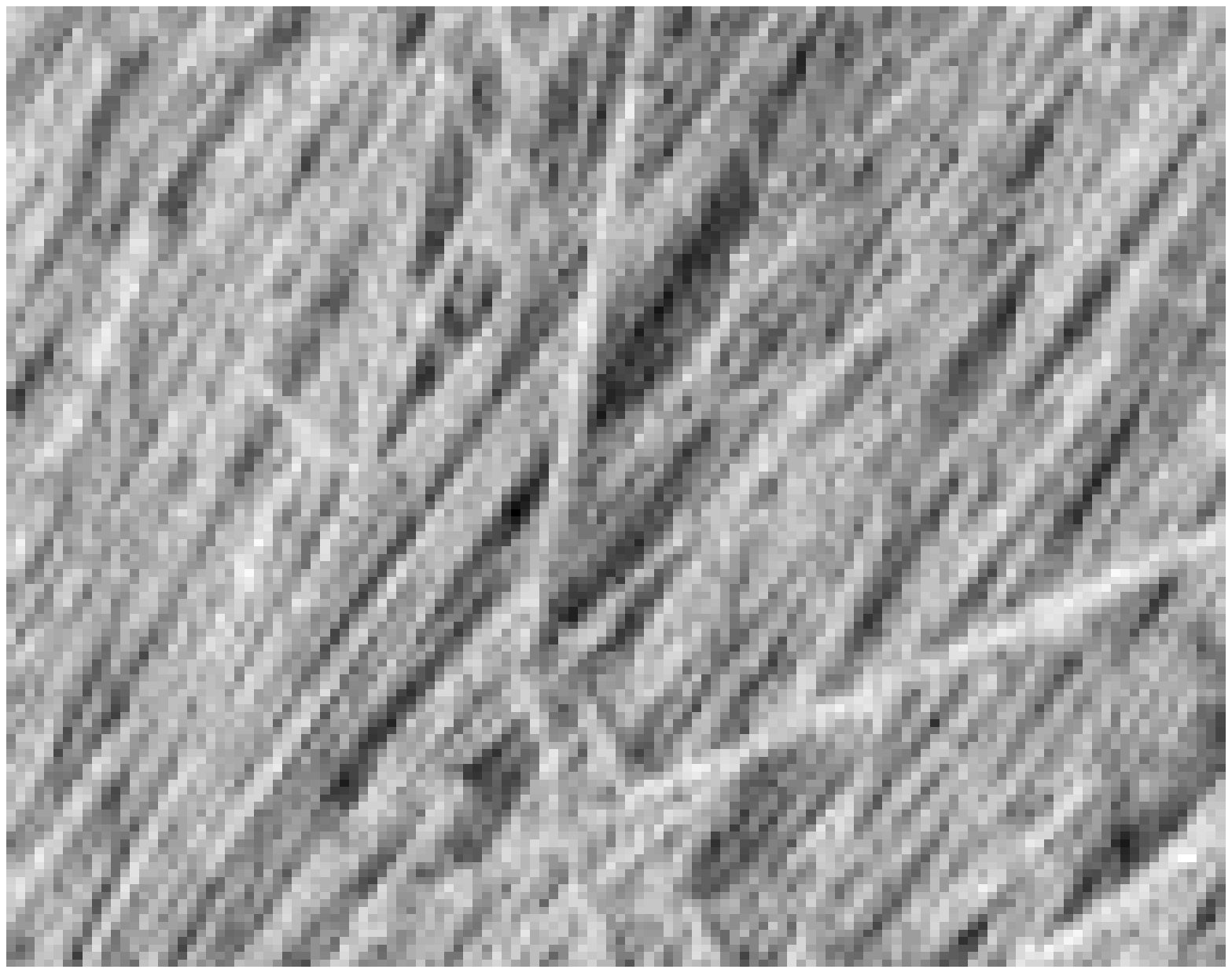}
\end{tabular}
\caption{Original image (a), noisy image (b) and denoised images using the variational approach (c) and the proposed MMSE estimator (d).} \label{fig:experiment2_Lp}
\end{figure}

\hfill \\
\label{sec:exp2}
In this experiment, we are interested in recovering an image (the \textit{Straw} image of size $128\times 128$) from its noisy observation affected by a noise
$\vect{n}$ uniformly distributed over the centered $\ell_p$ ball of
radius $\eta$ when $p \in\{1,2,3\}$. The translation invariant wavelet transform was used as a frame decomposition with a Symmlet filter of length 8 over 3 resolution levels. The $\ell_p$ norm ($p \in\{1,2,3\}$) was used for $N(\cdot)$ in \eqref{eq:delta}. Figs.~\ref{fig:experiment2_Lp} (a) and \ref{fig:experiment2_Lp} (b) show the original and noisy images using a uniform noise over the 
$\ell_2$ ball of radius $1600$. Figs.~\ref{fig:experiment2_Lp} (c) and \ref{fig:experiment2_Lp} (d) illustrate the denoising strategies 
based on the variational approach and the MMSE estimator using frame coefficients sampled with our algorithm. 

Table~\ref{tab:snr_ssim_denoising2} illustrates the SNR and SSIM values for noisy and denoised images using the proposed MMSE estimator with uniformly distributed noise for different values of $p$ and $\eta$.

\begin{table}[!ht]
\centering \caption{SNR and SSIM values for the noisy and denoised images.}
\begin{tabular}{|l|c|c|c|c|c|c|}
\hline
\multicolumn{2}{|c|}{} &Noisy & Wiener & Variational & MCMC\\
\hline
$\eta =300000$&SNR (dB)& 15.56 & 16.42 & 16.67 & \textbf{18.11}\\
\cline{2-6}
$p=1$         &SSIM    & 0.719 & 0.705 & 0.730 & \textbf{0.755}\\
\hline
$\eta =3000$ &SNR (dB) & 16.46 & 17.03 & 17.84 & \textbf{19.02}\\
\cline{2-6}
$p=2$       &  SSIM    & 0.749 & 0.720 & 0.758 & \textbf{0.796}\\
\hline
$\eta =700$ & SNR (dB) & 16.14 & 17.05 & 17.65 & \textbf{19.29}\\
\cline{2-6}
$p=3$         & SSIM   & 0.734 & 0.720 & 0.671 & \textbf{0.771}\\
\hline
\end{tabular}
\label{tab:snr_ssim_denoising2}
\end{table}
\newpage
This second set of image denoising experiments shows that the proposed approach performs well when using different kinds of frame representations and various noise properties.

\newpage


\newpage

\newpage

\newpage

\section{Conclusion}
\label{sec:conclusion}
This paper proposed a hierarchical Bayesian algorithm for frame coefficient from a noisy observation of a signal or image of interest.
The signal perturbation was modelled by introducing a bound on a distance
between the signal and its observation. A
hierarchical model based on this maximum distance property was then
defined. This model assumed GG priors for the
frame coefficients. Vague priors were assigned to the
hyper-parameters associated with the frame coefficient priors.
Different sampling strategies were proposed to generate samples
distributed according to the joint distribution of the parameters
and hyper-parameters of the resulting Bayesian model. The generated
samples were finally used for estimation purposes. Our validation
experiments showed that the proposed algorithms
provide an accurate estimation of the frame coefficients and
hyper-parameters. The good quality of the estimates was confirmed on
statistical processing problems in image denoising with multivariate noise uniformly distributed on some given ball. Despite its interest in dealing with bounded errors, this model was 
fewly investigated in the wavelet denoising literature.\\
The hierarchical model studied in this paper
assumed GG priors for the frame coefficients.
However, the proposed algorithm might be generalized to other
classes of prior models. Another direction of research for future
work would be to extend the proposed framework to situations where
the observed signal is degraded by a linear operator (e.g. blur operator).

\appendices
\section{Sampling on the unit $\ell_p$ ball}\label{append:a1}

This appendix explains how to sample vectors in the unit $\ell_p$
ball ($p \in ]0,+\infty]$) of $\RR^L$. First, it is interesting to
note that sampling on the unit ball can be easily performed in the
particular case $p = +\infty$, by sampling independently along each
space coordinate according to a distribution on the interval
$[-1,1]$. Thus, this appendix focuses on the more difficult problem
associated with a finite value of $p$. In the following, $\|\cdot\|_p$
denotes the $\ell^p$ norm. We recall the following theorem:
\begin{theorem}\rm{\cite{Song_97}} \hfill \\
\label{th:1} Let $\vect{A}=[A_1,\ldots,A_{L'}]^{\trans}$ be the
random vector of i.i.d. components  which have the following
$\mathrm{GG}(p^{1/p},p)$ pdf
\begin{equation}
f(a_i)=\frac{p^{1-1/p}}{2\Gamma(1/p)} \exp \left(-\frac{|a_i|^p}{p}
\right), \qquad a_i \in \RR.
\end{equation}
Let $\vect{U}=[U_1,\ldots,U_{L'}]^{\trans} =
\vect{A}/\|\vect{A}\|_p$. Then, the random vector $\vect{U}$ is
uniformly distributed on the surface of the $\ell_p$ unit sphere of
$\RR^{L'}$ and the joint pdf of $U_1,\ldots,U_{L'-1}$ is
\begin{equation}
f(u_1,\ldots,u_{L'-1})=\frac{p^{L'-1}\Gamma(L'/p)}{2^{L'-1}(\Gamma(1/p))^{L'}}
\left( 1-\sum_{k=1}^{L'-1}|u_k|^p \right)^{(1-p)/p}
\mathsf{1}_{D_{p,L'}}(u_1,...,u_{L'-1})
\end{equation}
where $D_{p,L'}= \{(u_1,...,u_{L'-1}) \in \RR^{L'-1} \mid
\sum_{k=1}^{L'-1}|u_k|^p < 1  \}$.
\end{theorem}
The uniform distribution on the unit $\ell_p$ sphere of $\RR^{L'}$
will be denoted by $\mathcal{U}(L',p)$. The construction of a random
vector distributed within the $\ell_p$ ball of $\RR^L$ with $L< L'$
can be derived from Theorem~\ref{th:1} as expressed below:
\begin{theorem}\rm{\cite{Song_97}} \hfill \\
\label{th:2} Let $\vect{U}=[U_1,\ldots,U_{L'}]^{\trans} \sim
\mathcal{U}(L',p)$. For every $L \in \{1,\ldots,L'-1\}$, the pdf of
$\vect{V}=[U_1,\ldots,U_L]^{\trans}$ is given by
\begin{equation}
q_1(u_1,\cdots,u_L) =
\frac{p^L\Gamma(L'/p)}{2^L(\Gamma(1/p))^L\Gamma((L'-L)/p)}
\left(1-\sum_{k=1}^{L}|u_k|^p
\right)^{(L'-L)/p-1}\mathsf{1}_{D_{p,L+1}}(u_1,...,u_L).
\end{equation}
\end{theorem}
In particular, if $p\in \NN^*$ and $L' = L+p$, we obtain the uniform
distribution on the unit $\ell_p$ ball of $\RR^L$.\\
Sampling on an $\ell_p$ ball of radius $\eta > 0$ is straightforwardly deduced
by scaling $\vect{V}$.

\bibliographystyle{IEEEbib}
\bibliography{biblio_chaari}

\end{document}